\documentclass[12pt]{article}
\usepackage{graphicx}
\newcommand{\ccdot}{ \! \cdot \!}
\usepackage{epsfig}

\begin{document}

\title{{\bf  RQM description of the charge form factor \\ 
 of the pion and its asymptotic behavior} } 
\author{ B.  Desplanques\thanks{{\it E-mail address:}  
desplanq@lpsc.in2p3.fr}  \\  
Laboratoire de Physique Subatomique et de Cosmologie \\
(UMR CNRS/IN2P3 -- UJF -- INPG) \\ 
  F-38026 Grenoble Cedex, France }

\sloppy

\maketitle

\begin{abstract}
\small{
The pion charge and scalar form factors, $F_1(Q^2)$ and $F_0(Q^2)$, 
are first calculated in different forms of relativistic quantum mechanics. 
This is done using the solution of a mass operator that contains both 
confinement and one-gluon-exchange interactions. Results of calculations, 
based on a one-body current, are compared to experiment for the first one. 
As it could be expected, those point-form, and instant and front-form ones 
in a parallel momentum configuration fail to reproduce experiment. The other results 
corresponding to a perpendicular momentum configuration (instant form in the Breit 
frame and front form with $q^+=0$) do much better. The comparison of charge
and scalar form factors shows that the spin-1/2 nature of the constituents plays an
important role. Taking into account that only the last set of results 
represents a reasonable basis for improving the description of the charge 
form factor, this one is then discussed with regard to the asymptotic QCD-power-law 
behavior $Q^{-2}$. The contribution of two-body currents in achieving 
the right power law  is considered while the scalar 
form factor, $F_0(Q^2)$, is shown to have the right power-law behavior in any
case. The low-$Q^2$ behavior of the charge form factor and the pion-decay
constant are also discussed.}
\end{abstract} 


\section{Introduction}
The pion has a double interest. As a physical system, it allows one 
to learn about its hadronic structure and how QCD is realized. 
As a test case, it can be used  
to check methods to describe properties of strongly interacting systems. 
Its charge form factor, which has been measured up to  
$Q^2\simeq 10$ (GeV/c)$^2$
\cite{Bebek:1978pe,Amendolia:1986wj,Volmer:2000ek,Horn:2006,Tadevosyan:2007,Huber:2007}, 
has thus been the object of a lot of attention. 
Noticing that the distinction is not always relevant as soon as some
approximations are made, it has  been considered in both field-theory 
\cite{Roberts:1996hh,Maris:1998,Maris:2000sk,Merten:2002nz,Simula:2002vm,Bakker:2001pk,deMelo:2002yq}
and relativistic-quantum-mechanics (RQM) approaches 
\cite{Isgur:1984jm,Chung:1988mu,Cardarelli:1994,Cardarelli:1995dc,Choi:1997iq,deMelo:1997cb,Allen:1998hb,Krutov:2002nu,Amghar:2003tx,Coester:2004qu,He:2004ba,Li:2008}.
Most often, results have been obtained and analyzed within 
the standard light-front  formalism ($q^+=0$). 

Recently, more extensive studies looking 
at the role of various momentum configurations within field-theory approaches, 
with the transfer momentum parallel or perpendicular either to the 
average momentum of the system or to the front-orientation, have been made 
\cite{Simula:2002vm,Bakker:2001pk,deMelo:2002yq}. A large sensitivity 
to various one and two-body contributions was observed  though 
the total result is well determined. Not much attention was however given 
to these results, perhaps because the momentum configuration $q^+=0$ 
generally imposes itself as the most appropriate. 

The situation in relativistic quantum mechanics is somewhat different. 
On the one hand, in absence of a close relationship to field theory, 
there are currently discussions as to which wave function or/and 
which current will be the best to reproduce experimental data 
\cite{Chung:1988mu,Allen:1998hb,Coester:2004qu,He:2004ba}. 
The only study that tried to incorporate a more founded wave function 
\cite{Cardarelli:1994,Cardarelli:1995dc} failed to account for the measurements 
of the pion charge form factor and, in this order, was forced
to invoke quark form factors. These ones could account for some missing 
physics which is expected to play some role like that one underlying 
the vector-meson-dominance phenomenology. On the other hand, it appeared 
that the estimate of this observable \cite{Amghar:2003tx,He:2004ba} 
could be very sensitive to the form employed in implementing 
relativity \cite{Dirac:1949cp}, quite similarly to what was observed for
spinless constituents \cite{Amghar:2002jx}. Moreover, an extension 
of this last work \cite{Desplanques:2004sp} showed a strong sensitivity 
of instant and front-form results to the momentum configuration, 
like in field-theory based approaches 
\cite{Simula:2002vm,Bakker:2001pk,deMelo:2002yq}. It would not be 
surprising whether a similar effect shows up in the pion case too. 
Altogether, there are therefore many reasons for motivating further 
studies of the pion charge form factor within relativistic quantum 
mechanics. One could add that specific predictions from QCD 
relative to the asymptotic pion charge form factor 
\cite{Brodsky:1975vy,Farrar:1979,Lepage:1979zb} 
or to the Goldstone nature of the pion have been hardly discussed  
within the RQM framework.

When looking at the pion form factor within relativistic quantum mechanics 
approaches, further work could be motivated by the numerous differences 
that are expected with the scalar-particle case considered in  previous works 
\cite{Desplanques:2001zw,Amghar:2002jx,Desplanques:2004sp}. 
Apart from the fact that the relevance for a physical system of conclusions 
achieved for an academic one has been questioned, the spin-1/2 nature 
of the constituents represents an important if not a major difference 
as will be seen. This can affect the solution of a mass operator, 
the current (Lorentz vector or Lorentz scalar) and, ultimately, the charge 
and Lorentz-scalar form factors as well as their asymptotic behavior. 
Confinement represents another difference characterizing hadronic 
physics. How these differences show up when comparing form factors
calculated in different forms (and different momentum configurations) together 
with a single-particle current is to be determined. This is our first goal. 

Anticipating that  a reasonable starting point for further improvement 
is only given by the standard instant- and front-form approaches
(Breit-frame and $q^+=0$ respectively), 
a second goal is to determine the role of two-body currents in getting 
the right $Q^{-2}$ QCD asymptotic behavior for the pion charge form factor 
within these approaches \cite{Farrar:1979,Lepage:1979zb}. 
Addressing this question in a RQM 
framework may look premature. Apart from the fact it should be ultimately 
considered, we notice that previous work with scalar particles 
succeeded relatively easily to reproduce the expected power-law 
behavior of form factors in the standard instant- and front-form 
approaches \cite{Amghar:2002jx,Desplanques:2004sp}. This gives us some
confidence in considering the problem here. It involves the interaction 
at short distances, due to one-gluon exchange and, of course, 
the spin of the constituents. With this respect and though there is 
no known scalar probe of practical relevance, the consideration 
of the scalar form factor, beside the charge one, is especially 
useful in revealing features specific to their asymptotic behavior. 
The contribution of two-body currents was considered in the past 
with the aim to restore the appropriate asymptotic behavior 
but this was for a calculation involving the ``point form" approach 
and scalar constituents \cite{Desplanques:2003}. Moreover, the origin 
of a wrong asymptotic behavior in this last approach differs 
from the one considered here for the pion in the standard instant- 
and front-form  approaches. 

While considering the two goals mentioned above, we found that the main 
trends evidenced by the results were rather insensitive to various ingredients 
entering the mass ope\-rator. Quantitatively, some sensitivity was observed 
however. We will therefore present here the main lines, leaving for future work 
a more complete discussion of the quantitative aspects. With this last respect, 
we will only mention the points that could be ques\-tio\-nable. Preliminary results 
were presented in ref. \cite{Desplanques:2004re} while the role of different 
forms was recently discussed by He {\it et al.} \cite{He:2004ba}, 
using phenomenological wave functions.

The plan of the paper is as follows. In the second section, we consider the
description of the pion in different forms. We, in particular, discuss 
the mass operator which contains both a confining part and a Coulomb one. 
The expression of the charge and scalar form factors in different forms is 
given in the third section. One-body and some two-body currents are considered.
The expression of the pion-decay constant, $f_{\pi}$, is also discussed as well
as quantities that involve this constant, namely the charge form factor
in the asymptotic domain and the charge radius (with some approximation). 
The fourth section is devoted to the presentation of the results 
with the one-body current and to their discussion, for the low- 
as well as the high-$Q^2$ range. The analysis of the asymptotic form factor 
is considered in the fifth section. The conclusion and further discussion 
is given in the sixth section. Three appendices contain details about 
accounting for the spin-1/2 nature of the constituents in the mass operator, 
the behavior of the solution at short distances for a strong Coulomb-type
interaction, and about two-body currents contributing to the charge form factor.

\section{Pion description in different forms}
Prior to any calculation of form factors in relativistic quantum 
mechanics, two ingredients have to be specified. On the one hand, 
a mass operator, whose solutions could be used in 
different forms, is needed. On the other hand, the relation of the momenta 
of the constituent particles to the total momentum, which depends on the form, 
is required. They are successively discussed in this section. 
\subsection{Mass operator}
 We here follow a previous work based on a quadratic mass 
operator together with a single-meson-exchange type interaction 
\cite{Amghar:2002jx}. This approach, 
specialized to scalar particles, was able to provide a good account of form 
factors corresponding to an ``exact'' theoretical model in both the instant- 
and front-form approaches. In the case of an infinite-mass 
exchange interaction, the ``exact'' form factor could be reproduced, ensuring
the correctness of a minimal number of ingredients for both the mass operator
and the currents. In the pion case, two further ingredients have 
to be considered: the spin-1/2 nature of the constituents and confinement. 
A mass operator that accounts for these features is the following:
\begin{eqnarray}
M^2_{\pi}\;\phi_0(k)&=&
\Big((2\,e_k+\sigma_{st}\;r-V_D)^2+ \Delta \Big)\;\phi_0(k) 
\nonumber \\  &&
-\int \; \frac{d\vec{k}'}{(2\,\pi)^3} \;
\frac{4\;g_{eff}^2\;\frac{4}{3}\;e_k\;e_{k'}
\Big(2- \frac{m^2_q\,(e^2_k+e^2_{k'})}{2\,e^2_k\;e^2_{k'}}\Big)}{
\sqrt{e_k}\;\;(\vec{k}-\vec{k}')^2\;\;\sqrt{e_{k'}}}\;\phi_0(k'), 
\label{x2a}
\end{eqnarray}
where $e_k =\sqrt{m_q^2+k^2}$. The spinor part of the pion wave function 
factors out and has therefore been omitted in writing the above equation 
(see appendix \ref{appa} for some detail). The only but important effect due 
to the spin-1/2 nature of the constituents, in compa\-rison with 
the scalar-particle case, is the replacement in the last term of a factor $m^2$ by 
$e_k\;e_{k'}\;\Big(2- m^2_q\,(e^2_k+e^2_{k'})/(2\,e^2_k\;e^2_{k'})\Big) $. 
The equation being written in momentum space, the distance $r$ should 
be considered as an operator. In order to simplify the writing of some 
expressions later on, we also introduce a wave function defined differently, 
$\tilde{\phi}(k)=\sqrt{e_k}\;\phi_0(k)$.

The confinement term together with the kinetic-energy one appears 
in a quadratic form, in accordance with the quadratic character 
of the mass operator. This model \cite{Basdevant:1986} can roughly 
reproduce the Regge trajectories and  the first radial excitations 
of mesons with a string tension $\sigma_{st} = 0.2\;$GeV$^2$ 
and $V_D = 0.486$ GeV.  The other part of the interaction is inspired from 
the single-gluon exchange one. It incorporates the standard Coulomb part 
but it also involves non-relativistic corrections that are usually 
omitted (see appendix \ref{appa}). It contains the QCD coupling, 
here replaced by an effective one, $g_{eff}^2/4$, 
and a factor $16/3$ representing the value taken by the color 
matrices for a meson. In accordance with the way we derive this part 
of the interaction, there is no specific contribution due to the standard 
spin-spin term. This one is actually cancelled in the present case 
by a term with the same spatial structure, generally omitted. 
The above effective one-gluon exchange interaction could {\it a priori} 
have a more complicated expression.  In absence of a detailed study, 
we assumed the simplest choice compatible with reproducing the one-gluon 
exchange in the perturbative regime, the only parameter being 
the effective coupling, $g_{eff}^2/4$. 
Once the above ingredients are fixed, the pion mass 
can be reproduced by choosing appropriately the quantity, $\Delta$. 
The equation can be solved in various ways. Working here in momentum space, 
we consider the operators $\sigma_{st}\;r$ and $(\sigma_{st}\;r)^2$ 
as the limit for $G$ going to infinity of the quantities, 
$G\;(1-{\rm exp}(-\sigma_{st}\;r/G))$ 
and $G^2\;(1-{\rm exp}(-(\sigma_{st}\;r/G)^2))$, 
and Fourier transform them. The solution is expected to have the 
high-momentum behavior, $\phi_0(k)\propto k^{-7/2}$, up to log corrections. 
For a strong enough coupling, these corrections can lead to a change 
of the power-law behavior. One thus expect $\phi_0(k)\propto k^{-3}$ for 
$(8/3)\;g_{eff}^2/(4\,\pi)=0.5$ (see appendix \ref{appb}). A critical regime 
could be reached when this quantity takes the value $2/\pi$ \cite{Leyouanc:1997}, 
which is in the range of expectations when the current QCD coupling, 
$g_{eff}^2/(4\,\pi)=\alpha_s$, is used. 
The power-law behavior of $\phi_0(k)$ is quite important as it determines 
that one for form factors \cite{Alabiso:1974sg}, unless some specific 
suppression occurs in relation with a particular probe. 

For practical purposes, we use $m_q=0.2\;$GeV, $M_{\pi}=0.14\;$GeV, 
$\sigma_{st}=1\;$GeV/fm$=0.2\;$GeV$^2$ and $V_D=0.5$ GeV. 
For  the strong QCD coupling, we distinguish two regimes, 
a low- and a high-energy one. In the first case, the effective coupling  
is taken as  $ (8/3)\;g_{eff}^2/(4\,\pi)=0.5$, which
corresponds to the case mentioned above where the wave function 
should scale like $ k^{-3}$ at large $k$. This coupling value is somewhat 
a compromise between  larger and smaller values expected respectively in
the low- and high-energy regime (see also below for other effects). 
In the second case, anticipating on the fact that the coupling 
should go to 0, it was assumed that the solution behaves like $ k^{-7/2}$ 
for $k \geq $5.6 GeV. As a result of these choices,  
the behavior of our solution slowly changes from the behavior $ k^{-3}$ 
obtained around 1 GeV to the one ascribed beyond 5.6 GeV. 
One then obtains the value of the last parameter, $\Delta=-0.487$ GeV$^2$.

Being interested here in {\it gross} features, we only considered the most 
important parts of the interaction: the confinement and the one-gluon 
exchange, which are essential to reproduce respectively the dominant aspects 
of the hadron spectroscopy and the asymptotic behavior of form factors. 
We did not therefore try to improve upon the above model. On the one hand, 
the derivation of the effective interaction to be used 
in relativistic quantum mechanics is largely open. This has been done 
to some extent in the scalar-particle case where the effect of retardation
was found to lead to an effective coupling 2-3 times smaller than the 
free one \cite {Amghar:2000pc}. The estimate of this effect and other ones 
is likely to be more complicated for the one-gluon exchange interaction \cite {Amghar:2000}, 
of interest here. On the other hand,  the running character of the strong
QCD coupling, $\alpha_s$, should be accounted for. Due to the relatively 
enhanced weight of high momenta in the solutions of eq. (\ref{x2a}), 
one can anticipate that  values of this coupling smaller than those currently
referred to for the low-energy domain should be used.
An important point is that, in both cases, the most singular part of the
interaction, which determines the asymptotic behavior of form factors 
\cite{Desplanques:2000ev} and is given  in eq. (\ref{x2a}) by the Coulomb term, 
be preserved (up to log terms). 
Of course, the interaction can be improved on a phenomenological basis, 
by requiring for instance a closer fit of the parameters to the pion 
radial excitations at 1300 MeV and 1800 MeV or to the pion-decay 
constant, $f_{\pi}$.
\subsection{Constituent and total momenta in different forms: relation}
In order to determine the expression of form factors, beside a solution 
of a mass operator discussed above, the relation of the constituent 
momenta, $\vec{p}_1, \;\vec{p}_2$,  to the total momentum, $\vec{P}$, 
and the internal variable, $\vec{k}$, is needed.  Two ingredients 
enter their determination. The first one involves the relation of the sum 
of the constituent 3-momenta and the total momentum: 
\begin{equation}
\vec{p}_1+\vec{p}_2-\vec{P}=
 \frac{ \vec{\xi } }{ \xi^0}\;(e_1+e_2-E_P),\label{moment}
\end{equation}
where  $\xi^{\mu}$ characterizes the approach under consideration and, 
especially, the symmetry properties of the hypersurface which physics 
is described on. Most often,  $\xi^{\mu}$ represents the orientation 
of a hyperplane and the above equation can then be obtained, up to a phase,  
by integrating plane waves, ${\rm exp}(i\,(p_1+p_2-P)\cdot x)$, 
over the hypersurface $\xi \cdot x=0$. The second ingredient involves 
a Lorentz-type transformation of the constituent momenta which generalizes 
that one introduced by Bakamjian and Thomas \cite{Bakamjian:1953kh}: 
\begin{equation}
\vec{p}_{1,2}=\pm\vec{k}
\pm\vec{w} \;\frac{\vec{w} \cdot \vec{k}}{w^0+1}
+\vec{w}\;e_k\;, \;\;\;\;
e_{1,2}=w^0\;e_k 
\pm\vec{w} \cdot \vec{k}\;, \label{y3b}
\end{equation}
where $w^0=\sqrt{1+\vec{w}\,^2}\;(w^2=(w^0)^2-\vec{w}\,^2=1 )$.  
Combining the two equations allows one to write:
\begin{equation}
 w^{\mu} =\frac{ P^{\mu}}{ 2\,e_k }
+ \frac{ \xi^{\mu} }{ 2\,e_k } \; \frac{4\,e_k^2-M^2}{
 \sqrt{ (\xi \cdot P)^2 + (4\,e_k^2-M^2)\;\xi^2 } +  \xi \cdot P} \, ,
\label{boost2} 
\end{equation}
where $w^{\mu}$ depends on the approach through that one of $\xi^{\mu}$.  
It is noticed that, $\xi^{\mu}$ representing an orientation, its scale 
is irrelevant, what can be checked on eq. (\ref{moment}) or on the 
last expression for $w^{\mu}$. The expressions taken by $\xi^{\mu}$ 
and $w^{\mu}$ are given in the following for different forms.

\noindent
$\bullet$ {\it Instant-form approach} \\ 
\begin{equation}
\xi^{\mu}\propto\lambda_0^{\mu}\;, 
\hspace{1cm} {\rm with}\;\; 
(\lambda_0^0,\;\vec{\lambda}_0)=(1,\; 0) \; (\lambda^2=1). 
\label{y3d}
\end{equation}
\begin{equation}
\vec{w}= \frac{\vec{P}}{2\,e_k}, 
\hspace{1cm}
w^0= \frac{\sqrt{4\,e_k^2+P^2}}{2\,e_k}\;.  \label{y3c}
\end{equation}

\noindent
$\bullet$ {\it Front-form approach} \\ 
\begin{equation}
\xi^{\mu} \propto \omega^{\mu}\;, \
\hspace{1cm} {\rm with}\;\; 
(\omega^0,\;\vec{\omega}) \propto (1,\;\hat{n}), \;(\omega^2=0)\,,\label{y3f}
\end{equation}
\begin{equation}
 \vec{w} =\frac{ \vec{P} }{ 2\,e_k }
+ \frac{ \hat{n} }{ 4\,e_k } \; \frac{ 4\,e_k^2-M^2 }{E_P-\vec{P}\cdot 
\hat{n}}\;,  
\hspace{1cm}
w^0 = \frac{E_P }{ 2\,e_k } 
+\frac{4\,e_k^2-M^2 }{4\,e_k\;(E_P-\vec{P}\cdot \hat{n}) }\;. \label{y3e}
\end{equation}
where $\hat{n}$ has a fixed orientation. 

\noindent
$\bullet$ {\it Dirac inspired point-form approach} 
\cite{Desplanques:2004rd} \\ 
\begin{equation}
\xi^{\mu}\propto u^{\mu}\;, \
\hspace{1cm} {\rm with}\;\; 
(u^0,\;\vec{u}) \propto (1,\;\hat{u}), \;(
u^2=0)\,, \label{y3m}
\end{equation}
\begin{equation}
 \vec{w} =\frac{ \vec{P} }{ 2\,e_k }
+ \frac{ \hat{u} }{ 4\,e_k } \; \frac{ 4\,e_k^2-M^2 }{E_P-\vec{P}\cdot 
\hat{u}}\;,  
\hspace{1cm}
w^0 = \frac{E_P }{ 2\,e_k } 
+\frac{4\,e_k^2-M^2 }{4\,e_k\;(E_P-\vec{P}\cdot \hat{u}) }\;. \label{y3l}
\end{equation}
where $\hat{u}$, contrary to the front-form case, has no fixed orientation.

In the absence of interaction ($2\,e_k=M$), the three expressions 
obtained in different forms for $ \vec{w}  $ and $w^0 $ become identical 
to  the standard kinematical boost for free particles. In this limit, 
one recovers the expressions relevant to an earlier ``point-form'' approach
 \cite{Bakamjian:1961,Sokolov:1985jv,Klink:1998}:
\begin{equation}
\xi^{\mu} \propto \lambda^{\mu}\;, 
\hspace{1cm} {\rm with}\;\; \lambda^{\mu}=P^{\mu}/\sqrt{P^2}.
\label{y3i}
\end{equation}
\begin{equation}
\vec{w} =\frac{ \vec{P} }{M}\;,
\hspace {1cm}
w^0 = \frac{E_P }{ M }. \label{y3h}
\end{equation}
This ``point-form'' resembles 
the Dirac point form in that the dynamical or kinematical
character of the Poincar\'e generators is the same. However, 
as mentioned elusively by Bakamjian \cite{Bakamjian:1961} 
and  explicitly by Sokolov \cite{Sokolov:1985jv}, 
this ``point-form'' implies describing the physics on hyperplanes 
perpendicular to the velocity of the system and differs from the 
Dirac one, which involves a hyperboloid surface. The kinematical 
character of boosts and rotations in the above 
``point-form'' stems from the invariance under Lorentz transformations 
of the hyperplane defined by the condition that $v \cdot x$ is a constant, 
where $v$ is the 4-velocity of the system ($v^{\mu} \propto P^{\mu}$). 
To some extent, the property 
is trivial as the frame used to describe the system changes 
at the same time as this one is boosted. A related 
feature is the fact that it requires a  constraint on the interaction 
much weaker than in the other approaches. When deriving a mass 
operator, any Lorentz invariant interaction is sufficient while, in 
the other cases, the consideration of the interaction at all orders 
is required \cite{Desplanques:2004rd}. A similar statement holds for other
quantities such as the charge form factor at $Q^2=0$ or the pion-decay 
constant (see below). The above ``point-form'' approach 
thus possesses properties that make it definitively different from 
the other ones. In some sense, it contains the effect of the boost 
transformation common to all approaches but without the constraints 
attached to the fact that the hypersurface which the system is described on
is uniquely defined and independent of its velocity. As is well known, 
this requires interaction effects, represented here by the  $\xi^{\mu}$ 
dependent term at the r.h.s. of eq. (\ref{boost2}).

\section{Current and form factors}
\begin{figure}[tb]
\begin{center}
\includegraphics[width=10cm]{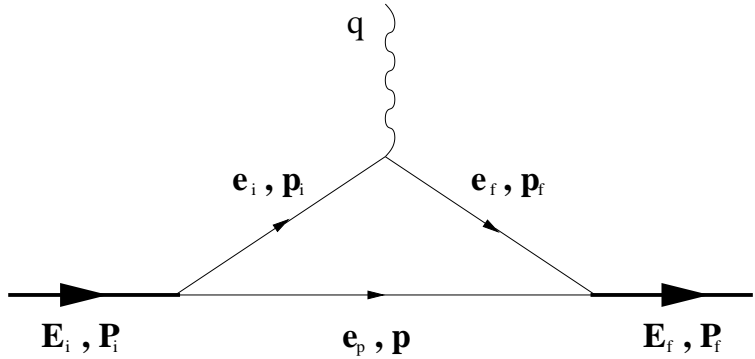}
\end{center}
\caption{Graphical representation of a virtual photon absorption on a pion  
together with kinematical definitions.}\label{fig:1}
\end{figure} 
We consider in this section the pion form factors. There are two of them,
corresponding to the coupling to a photon ($F_1(Q^2)$: charge form factor) and 
to a possible scalar probe ($F_0(Q^2)$: scalar form factor). Their definition 
we refer to here may be found in ref. \cite{Amghar:2002jx}. 
Considering both of them can
give a better insight on the implementation of relativity, especially 
with respect to the non-zero spin of the constituents. The studies 
are performed on the basis of a single-particle current. Two-body currents 
will be considered but only for those approaches that already provide 
a relatively good account of experiment. We therefore discard large
contributions from two-body currents that could be required 
to preserve properties related to the Poincar\'e space-time translation 
invariance \cite{Desplanques:2004sp,Desplanques:2008}. In short, 
these last currents account in a RQM framework for the equality 
of the momentum  that is separately transferred to the constituents 
and to the whole system, which is fulfilled in field-theory approaches 
in a straightforward way. 
\subsection{One-particle current}
The form of the single-particle current to be used with the above mass operator
has been discussed at length in ref. \cite{Amghar:2002jx} for spinless 
constituents. Interestingly, it has been found, since then, that form factors 
in different forms, due to minimal consistency requirements, could be written 
in a unique way \cite{Desplanques:2004sp}. They should be completed to account
for the spin-1/2 nature of quarks. Referring to fig. \ref{fig:1} 
for the kinematical notations, they now read:
\begin{eqnarray}  
F^{(1)}_1(Q^2) \;\Big[ F^{(1)}_0(Q^2)   \Big]=   
\int \frac{ d\vec{p}}{(2\pi)^3} \;\;
\frac{\xi_f \ccdot (p_f+p)\;\;\xi_i \ccdot (p_i+p)}{
e_p\;(2\,\xi_f \ccdot p_f) \;(2\,\xi_i \ccdot p_i) }\;\;
\phi_0 \Big((\frac{p_f-p}{2})^2\Big) \; \phi_0 \Big( (\frac{p_i-p}{2})^2 \Big)  
  \nonumber  \\  \times 
\;\;\Big( (p_f+p)^2 \, (p_i+p)^2 \Big)^{-1/4}\;\;\;
\frac{ I\ccdot (\xi_f+\xi_i )}{
(p_f+p_i+2\,p) \ccdot  (\xi_f+\xi_i )}\;
\;\Big[\frac{S}{4\,m_q} \;\;cf_0 \Big]. 
  \label{y3a}
\end{eqnarray}  
In  this equation, $I^{\mu}$ and $S$ result from the following trace 
over $\gamma$ matrices:
\begin{eqnarray}
I^{\mu}&=&4\;{\rm Tr}\Big[ \gamma_5\; \frac{(m_q+\gamma \cdot p_i)}{2} 
\;\gamma^{\mu}\;\frac{(m_q+\gamma \cdot p_f)}{2}\;
\gamma_5\; \frac{(m_q-\gamma \cdot p)}{2}\Big],
\nonumber \\
&=&2\; \Big[p_i^{\mu}\;(p_f \cdot p)+ p_f^{\mu}\;(p_i \cdot p)- 
p^{\mu} \;(p_i \cdot p_f) + m_q^2\;(p_i^{\mu}+p_f^{\mu}+p^{\mu})\Big],
\nonumber \\
S&=&4\;{\rm Tr}\Big[ \gamma_5\; \frac{(m_q+\gamma \cdot p_i)}{2} \;
 \frac{(m_q+\gamma \cdot p_f)}{2}\;\gamma_5\; \frac{(m_q-\gamma \cdot
 p)}{2}\Big]\,,
\nonumber \\
&=&2\;m_q\; \Big[p_f \cdot p+p_i \cdot p+ p_i \cdot p_f + m_q^2\Big]\,.
\label{x3b} 
\end{eqnarray}
In the case of a point-form approach inspired from the Dirac one, 
it turns out that expressions for form factors can be obtained 
from the above ones in the front form 
by integrating them over the orientation of the front together with 
an appropriate weight. They can be obtained from a straightforward 
generalization of those for the scalar-particle case \cite{Desplanques:2004sp}. 

It can be checked that the expression of the charge form factor 
at $Q^2=0$ considerably simplifies. Making an appropriate change 
of variable, it reduces in all cases to a unique expression: 
\begin{equation}
F^{(1)}_1(Q^2=0)=\int \frac{ d\vec{k}}{(2\pi)^3 } \; \phi^2_0(k)=
 \frac{1}{(2\pi)^3}\int \frac{ d\vec{k}}{e_k } \; \tilde{\phi}^2(k)=1.
\end{equation}
We stress that the result is independent of both the momentum of the system, 
$\vec{P}$, and the orientation of the hyperplane given by $\xi^{\mu}$, 
a property which is not trivial in most cases. With this respect, 
the presence of the factor, $(p_f+p_i+2\,p) \ccdot \xi$, 
at the denominator in eq. (\ref{y3a}) is essential. This one 
accounts in a hidden way for some two-body currents  \cite{Amghar:2002jx}.
It is also noticed that the above expression of the charge form factor 
at $Q^2=0$ is fully consistent with the expression of the normalization 
of the solution, $\phi_0(k)$, of the mass operator, eq. (\ref{x2a}), 
which involves the same integrand. 

In the case of the scalar form factor, and for the instant form, 
it is generally found that its value at $Q^2=0$ depends on the
momentum of the system. In other words, it does not verify a minimal Lorentz
invariance property. This one can be fulfilled by introducing a correction
factor in the expression of the scalar form factor, $F_0(Q^2)$. 
Denoted $ cf_0$, this quantity amounts to account for two-body currents. 
Its expression, whose form is suggested by the consideration of a simple
interaction model, has been considered independently \cite{Desplanques:2004sp}.
Symmetrizing its effect between initial and final states, it is taken as:
\begin{equation}
cf_0=1+\frac{1}{2}\;(g(k_i)+g(k_f))\;e_p \;
\frac{(e_f+e_p)(e_i+e_p)-E_f\;E_i}{(e_f+e_p)(e_i+e_p)(e_f+e_i)},
\end{equation}
where the function $g(k)$ can be obtained from the following quadrature:
\begin{equation} 
\frac{g(k)}{8}\;e_k\;(4\,e_k^2-M^2)\;\phi^2_0(k)=
\int^{\infty}_k dk'\;  k'\; e_{k'} \; \phi^2_0(k')\,.
\label{x3bb}
\end{equation}
In the case of a scalar-particle model 
together with a zero-range interaction, $g(k)$ 
is equal to 1 and the above factor $cf_0$ then allows one 
to reproduce the ``exact'' scalar form factor, $F_0(Q^2)$.

A few remarks are in order here. A first one concerns the Lorentz-invariance 
properties of form factors in different forms. 
While the point-form approaches evidence this property, 
the instant- and front-forms ones do not. These last approaches 
can thus be considered for different momentum configurations which include 
the standard ones (Breit frame and $q^+=0$ respectively) but also 
non-standard momentum configurations like a parallel one where the transfer momentum 
is oriented along the average momentum of the system 
($(\vec{P}_i-\vec{P}_f) \parallel (\vec{P}_i+\vec{P}_f)\parallel \vec{n}; 
\;E_i \neq E_f$). 
Their comparison could be instructive but this supposes that no other symmetry 
is significantly broken at the same time, which has to be checked. 
A second remark concerns the expressions of form factors in the front-form 
case. The current ones, generally given in terms of the Bjorken variable, $x$, 
and the transverse momentum, are recovered from  eq. (\ref{y3a}) 
by making a change of variable. The last remark concerns the calculation 
of form factors in the earlier ``point-form''. As the initial and final states
have generally different momenta, two different 4-vectors, 
$\xi_i^{\mu}=P_i^{\mu}/M$ and $\xi_f^{\mu}=P_f^{\mu}/M$, and therefore 
two different hyperplanes, are then involved \cite{Desplanques:2001ze}.

When comparing the various approaches, 
an important feature emerges. The boost transformation allowing one 
to get the wave functions describing the initial or final states 
from the solution of a mass operator only depends on the
constituent mass in the standard instant- and front-form approaches 
\cite{Desplanques:2001ze,Coester:2004qu} while it also involves 
the mass of the system in all other cases. In a few of them considered 
later on (point-form and non-standard approaches in a parallel momentum 
configuration together with $|\vec{P}_i+\vec{P}_f| \rightarrow \infty$), 
form factors in the single-particle current approximation 
are shown to depend on the momentum transfer $Q$ 
through the ratio $Q/M$ \cite{Desplanques:2004sp}. This has striking 
consequences in the case where the mass of the system is small 
in comparison with the sum of the constituent ones, like for the pion. 
\subsection{The pion-decay constant, $f_{\pi}$}
At first sight, the calculation of the pion form factors can be performed 
independently of any knowledge about the pion-decay constant, $f_{\pi}$ 
($\simeq 93$ MeV in our conventions). 
This quantity is nevertheless relevant here as it enters in the QCD 
prediction of its asymptotic behavior for the charge one. It is successively
considered in the instant and front forms.

\noindent
$\bullet$ {\it Expression of $f_{\pi}$ in the instant form}\\
In a first approximation, the pion-decay constant in the instant form 
could be obtained from considering the matrix element of the axial current
between the pion and the vacuum:
\begin{eqnarray}
 f_{\pi}\; P_{\pi}^{\mu} =(?) \;\frac{\sqrt{3}}{(2\,\pi)^3}
 \int d\vec{p}_1 \;\;\Big(\frac{m_q\;}{e_k}\; \tilde{\phi}(k)\Big)\; \;m_q\;
  \frac{e_1+e_2}{2\,e_1\;e_2}\;
\nonumber \\ \times {\rm Tr} 
 \Big[\frac{ (m_q+\gamma \cdot p_1)\;\gamma^{\mu}\,\gamma_5\; 
 (m_q-\gamma \cdot p_2)\,\gamma_5 }{4\,m_q^2}\Big]\, ,
 \label{fpi1}
\end{eqnarray}
where the expression of $k$ is given at this point by $k^2=(p_1+p_2)^2/4-m_q^2$. 
Apart from the fact that the consideration of time and spatial components 
give different answers, which is not a surprise in a non-covariant approach 
(this is reminded by a question mark in front of the r.h.s. 
of the above equation), it is noticed that, in the c.m., the l.h.s. 
is proportional to the pion mass while the r.h.s. is not. 
A similar problem is encountered for the charge form factor,
revealing striking features for a strongly bound system \cite{Amghar:2002jx}.
It points to a missing contribution in the time component of the current 
given by the trace at the r.h.s. of eq. (\ref{fpi1}). This one, given 
by $(e_1+e_2)/m_q$, should be completed by the quantity $(E_{\pi}-e_1-e_2)/m_q$, 
which  can be seen to be an interaction term and provides a two-body current 
using eq. (\ref{x2a}). The expression of $f_{\pi}$ so obtained is independent 
of the component of the current and is given by:
\begin{equation}
f^{IF}_{\pi} 
=\frac{\sqrt{3}}{(2\,\pi)^3} \int 
\frac{d\vec{p}_1\;(e_1+e_2)}{2\,e_1\;e_2}\; 
\Big(\frac{m_q}{e_k}\; \tilde{\phi}(k)\Big)\,.
 \label{fpi2}
\end{equation}
By making a change of variable given by eqs. (\ref{y3b}, \ref{y3c}), 
it is found to be equal to: 
\begin{equation}
f^{IF}_{\pi} 
=\frac{\sqrt{3}}{(2\,\pi)^3} \int \frac{d\vec{k}}{e_k}\; 
\Big(\frac{m_q}{e_k}\;  \tilde{\phi}(k)\Big),
 \label{fpi4}
\end{equation}
which is independent of the momentum of the pion, $\vec{P}$, and therefore 
Lorentz invariant. 

\noindent
$\bullet$ {\it Expression of $f_{\pi}$ in the front form}\\
The expression of $f_{\pi}$ in the front form in terms of the Bjorken variable 
and the transverse momentum can be found in different works. 
With our convention for the wave function, $\phi_0(k)$, it reads: 

\begin{equation}
f^{FF}_{\pi}
=\frac{\sqrt{3}}{(2\,\pi)^3} \int d^2k_{\perp} \;
\frac{dx}{\sqrt{x\,(1-x)}} 
\frac{m_q}{\sqrt{m_q^2+k^2_{\perp}}}\;\tilde{\phi}(k),
 \label{fpi5}
\end{equation}
where the argument of the wave function, $k$, is defined 
by $k^2=(m_q^2+k^2_{\perp})/(4x(1-x))-m_q^2$. Not surprisingly, 
the above result can be recovered from the instant-form expression, 
eq. (\ref{fpi1}), in the limit $\vec{P} \rightarrow \infty$ and, of course, 
from eq. (\ref{fpi4}) by making the extra change of variable, 
$k^z=(1-2x)\sqrt{(m_q^2+k^2_{\perp})/(4x(1-x))}$. One has therefore:
\begin{equation}
f_{\pi}=f^{IF}_{\pi}=f^{FF}_{\pi}.
\end{equation}
\noindent
$\bullet$ {\it Expression of $f_{\pi}$ for an arbitrary hyperplane}\\
An expression of $f_{\pi}$, which is generalized to an arbitrary hyperplane 
of orientation $\xi^{\mu}$ and evidences the main ingredients, is given by:
\begin{eqnarray}
 f_{\pi}\; &= &\frac{\sqrt{3}}{(2\,\pi)^3}
  \int \frac{d\vec{p}_1}{e_1} \;\frac{d\vec{p}_2}{e_2}\;  m_q\; 
  \frac{\xi \cdot (p_1+p_2)}{2\,\xi^0 } \;
  \delta \Big(\vec{p}_1+\vec{p}_2-\vec{P}-
 \frac{ \vec{\xi } }{ \xi^0}\;(e_1+e_2-E_P)\Big)
\nonumber \\  && \hspace*{-5mm}\times \,
{\rm Tr} 
 \Big[\frac{  (m_q+\gamma \cdot p_1)\;\xi \cdot \gamma \gamma_5\; 
 (m_q-\gamma \cdot p_2)\gamma_5  }{4\,m_q^2}\Big]\frac{1}{\xi \cdot (p_1+p_2)}
\;\Big(\frac{m_q\;}{e_k}\;\tilde{\phi}(k)\Big)\,.
 \label{fpi6}
\end{eqnarray}
It simplifies to read:
\begin{equation}
 f_{\pi} = \frac{\sqrt{3}}{(2\,\pi)^3}
\int \frac{d\vec{p}_1}{e_1} \;\; \frac{\xi \cdot (p_1+p_2)}{2\,\xi \cdot p_2}\;\;
 \Big(\frac{m_q\;}{e_k} \;\tilde{\phi}(k)\Big) \, . 
\label{fpi7}
\end{equation}
The structure of this last expression is somewhat similar to that one used 
for the charge form factor, eq. (\ref{y3a}), at $Q^2=0$. Like for this one, 
it can be checked that the result evidences the property to be both 
independent of the velocity of the system (therefore Lorentz invariant) 
and of the orientation of the hyperplane, $\xi^{\mu}$. By performing 
a change of variable, it can be cast into the form of either eq. (\ref{fpi4}) 
or eq. (\ref{fpi5}) .

\subsection{Two-body currents}
\begin{figure}[tb]
\begin{center}
\includegraphics[width=12cm]{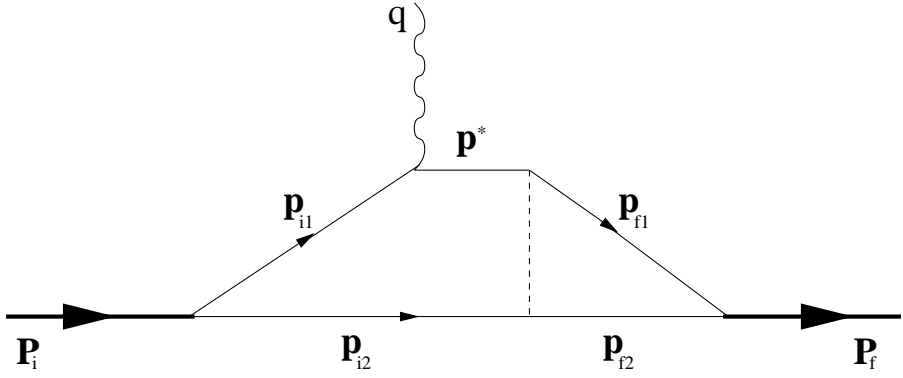}
\end{center}
\caption{Graphical description of a virtual-photon absorption on a pion  
with evidencing the contribution of one-gluon exchange representing 
the Born amplitude: kinematical definitions. A similar diagram with the gluon
exchange in the initial state should be considered.}\label{fig:2}
\end{figure} 
We here consider two-body currents that are required to recover the full Born
amplitude depicted in fig. \ref{fig:2}, taking into account that part of this
diagram with a positive-energy quark is already included in the contribution of
the single-particle current previously discussed, eq. (\ref{y3a}). This will be
done for the pion charge form factor where the problem of recovering the QCD
prediction for the asymptotic behavior \cite{Farrar:1979,Lepage:1979zb} 
is the most crucial. This one is given by:
\begin{equation}
F_1(Q^2)_{Q^2 \rightarrow \infty}= 16\,\pi\;f^2_{\pi} \; \frac{\alpha_s }{Q^2}.
\label{asymp}
\end{equation}
The problem is examined in both the standard instant and front forms, 
which are the only ones to provide a reasonable starting point for 
considering the contribution of interest in the present work. 
The expression of the two-body currents is given here in the simplest cases 
while the derivation is considered in appendix \ref{appc}. 
Two-body currents could also be considered for the scalar form factor but, as
this one turns out to have the right asymptotic power-law behavior, the necessity
for studying them is not as strong as for the charge form factor.

\noindent
$\bullet$ {\it Expression of $F^{(2)}_1(Q^2)$ in the instant form 
(standard Breit frame: $\vec{P}_f=-\vec{P}_i=\vec{Q}/2$)}\\
The  two-body contribution to the charge pion form factor in
the instant form may be expressed as a double integral over the momenta of that
quark in the initial or final states which does not interact directly with the
external probe (see fig. \ref{fig:2}). It reads:
\begin{eqnarray}
F^{(2)}_1(Q^2)&\!=\!& 
\int \frac{d\vec{p}_{i2}}{(2\,\pi)^3}\; 
\; \frac{e_{i1}\!+\!e_{i2}}{2\,e_{i1}\;e_{i2}}
\;\Big(\frac{m_q}{e_{k_i}}\; \tilde{\phi}(k_i) \Big) 
\;\int \frac{d\vec{p}_{f2}}{(2\,\pi)^3}\;
\;\frac{e_{f1}\!+\!e_{f2}}{2\,e_{f1}\;e_{f2}}
\;\Big(\frac{m_q}{e_{k_f}}\; \tilde{\phi}(k_f) \Big)
\; \hspace{1cm} 
\nonumber \\
&&\times \frac{4\pi\,\alpha_s\;\frac{4}{3}}{\mu^2-(p_{i2}\!-\!p_{f2})^2}\;
\Big( I^{-}_{ps}+I^{-}_{pv}+ I^{+}_{ps}+I^{+}_{pv}\Big)
+ \;({\rm  term:} \;i \leftrightarrow f) \, . 
\label{twobod1}
\end{eqnarray}
where quantities that could be related to the pion-decay constant, $f_{\pi}$, 
have been purposely factorized. The gluon propagator is made apparent. To keep
track of its effect in formal developments, we ascribe to the gluon a mass,
$\mu$, which is set to zero is actual calculations. 

The various quantities at the last line refer to an intermediate 
particle with negative and positive values of the quantity $p^{*0}$ 
and to the pseudo-scalar and pseudo-vector parts of the Fierz-transformed 
gluon-exchange interaction. They are respectively represented by superscripts 
$-$ and $+$  and subscripts $ps$ and $pv$.  As for the one-body part,
the matrix element of the current is divided by the quantity representing 
the sum of the kinetic energies of the constituents in the initial 
and final states, $e_{i1}+e_{f1}+e_{i2}+e_{f2}$. They also contain factors
relative to the intermediate quark propagator.  The summation over 
the quark spins has been performed. Their expressions read:
\begin{eqnarray}
&&I^{-}_{ps}=\frac{1}{(e_{i1}+e_{f1}+e_{i2}+e_{f2})\; \;  
2\,e^{*}_f\;(e^{*}_{f}+e_{i1})}\;
\nonumber \\ 
&& \times 2\,\frac{m_q^2\!+\!  p_{f1} \ccdot p_{f2}}{m_q^2}
\Bigg[  p_{i1} \ccdot p_{i2}\;\tilde{p}^{\,-\,0}_f
 - p_{i1}\ccdot \tilde{p}^{\,-}_f\;p_{i2}^{0} 
 + p_{i2}\ccdot \tilde{p}^{\,-}_f\;p_{i1}^{0} 
 +m_q^2\,(\tilde{p}^{\,-\,0}_f\! +\! p_{i1}^{0} \!+\!p_{i2}^{0} ) \Bigg]\, ,
\nonumber \\
&&I^{-}_{pv}=-\frac{1}{(e_{i1}+e_{f1}+e_{i2}+e_{f2})\; \;  
2\,e^{*}_f\;(e^{*}_{f}+e_{i1})}\;
\nonumber \\
&&\times 
\Bigg[\tilde{p}^{\,-}_f \ccdot (p_{f1}\!+\!p_{f2})\; (p_{i1}^{0} \!+\!p_{i2}^{0})
+(p_{i1}\!+\!p_{i2}) \ccdot (p_{f1}\!+\!p_{f2})\;\tilde{p}^{\,-\,0}_f
- \tilde{p}^{\,-}_f \ccdot (p_{i1}\!+\!p_{i2})\; (p_{f1}^{0} \!+\!p_{f2}^{0})
\nonumber \\
&&\;\;\;+p_{i2} \ccdot (p_{f1}\!+\!p_{f2})\;p^{0}_{i1}
-p_{i1} \ccdot (p_{f1}\!+\!p_{f2})\;p^{0}_{i2}
+(m_q^2+p_{i1}\ccdot p_{i2})\;(p_{f1}^{0} \!+\!p_{f2}^{0})
\Bigg] \, ,
\nonumber \\  \nonumber  \\
&&I^{+}_{ps}+I^{+}_{pv}= \frac{e^{*}_f-e_{i1}}{ 
(e_{i1}+e_{f1}+e_{i2}+e_{f2})\; \;  2\,e^{*}_f\;
(e_{i1}+e^{*}_f+2\,e_{i2})\;(e^{*}_f+e_{i2}+e_{f1}+e_{f2})}
\nonumber \\ && \times 
\Bigg[\Bigg( 2\,\frac{m_q^2\!+\!  p_{f1}\!\cdot\! p_{f2}}{m_q^2}
-\frac{(m_q^2\!+\!p_{i2}\ccdot\tilde{p}^{+}_f)) 
+(m_q^2\!+\!p_{f1} \ccdot p_{f2})}{m_q^2\!+\! \tilde{p}^{+}_f\ccdot p_{i2}} \Bigg)
\nonumber \\  && \hspace*{1cm} \times
\Big(   p_{i1} \ccdot p_{i2}\; \tilde{p}^{\,+\,0}_f
 - p_{i1}\ccdot \tilde{p}^{+}_f\; p_{i2}^0 
 + p_{i2} \ccdot \tilde{p}^{+}_f\; p_{i1}^0 
 +m_q^2\, (\tilde{p}^{\,+\,0}_f\! +\! p_{i1}^0 \!+\!p_{i2}^0 )\Big) \Bigg]\, ,
 \\
\nonumber \\
&& {\rm with}\; e^*_f=\sqrt{m_q^2+\vec{p}^{\,*2}_f}, \;\;
\tilde{p}^{\,\pm\,0}_f=\pm e^*_f, \;\;
\vec{\tilde{p}}^{\,\pm}_f= \vec{p}^{\,*}\, .\nonumber\\ \nonumber
\label{twobod2}
 \end{eqnarray}
The above appearance of contributions referring to the negative-energy 
component of the intermediate particle in fig. \ref{fig:2}, $I^{-}_{ps}$ 
and $I^{-}_{pv}$, is evident as they involve degrees of freedom 
that are not explicitly accounted for in a RQM approach. 
From their derivation, it can be checked that 
they vanish at zero momentum transfer. The separation into pseudo-scalar 
and pseudo-vector contributions is justified by the different role 
they play  as for the asymptotic behavior of the pion charge form factor
and, to some extent, for the charge radius.

The presence of contributions referring to the positive-energy component 
of the intermediate particle is less trivial. As the spin part 
of the pion wave function in the RQM formalism assumes a unique form, 
$\propto \bar{u}(p_{1}) \,i\, \gamma_5\, v(p_{2})$, it cannot 
fully account for that part of the Fierz-transformed interaction involving 
pseudo-vector currents. The discarded term  has an off-energy shell
character  however and, thus, can contribute to the form factor. 
The corresponding contribution involves both pseudo-scalar 
and pseudo-vector terms, $I^{+}_{ps}$ and $I^{+}_{pv}$. These ones do not
vanish at zero momentum transfer but their sum does as a result of a close 
relationship. For this reason, only their sum has been given in the above
expression of two-body currents. At non-zero momentum transfers, 
it turns out that the cancellation still holds but this result is specific 
to the Breit frame considered here ($\vec{P}_f=-\vec{P}_i=\vec{Q}/2$). An
expression valid for an arbitrary frame is given in the appendix \ref{appc}.

We stress that, apart from neglecting higher-order terms in the coupling, 
$\alpha_s$, the  above two-body contribution 
is entirely determined by  recovering the Born amplitude shown 
in fig. \ref{fig:2} after it is added to the single-particle contribution. 
Getting this  amplitude with the right strength therefore supposes 
that the contribution is correctly calculated 
as far as some minimal ingredients are concerned. 

From a rough examination of the above equations, one can convince oneself 
that the large $Q^2$ limit of $F^{(2)}_1(Q^2)$ in the Breit frame 
will take the expected form given by eq. (\ref{asymp}), except perhaps 
for some factor. This one requires some care, especially with 
the treatment of the components of the constituent momenta perpendicular 
to the momentum transfer, $\vec{Q}$. Contrary to a naive expectation, 
these ones can be shifted from their zero value by an amount 
approximately given by $Q\;(k/e_k)$. In order to get a correct expression 
of the asymptotic limit, it is appropriate to make a change of variable: 
$\vec{p}_1,\;\vec{p}_2 \rightarrow \vec{k}, \;\vec{P} $. One then obtains:
\begin{eqnarray}
F^{(2)}_1(Q^2)_{Q^2\rightarrow \infty}&=& 16\,\pi \;\frac{\alpha_s }{Q^2}\;\;
\frac{\sqrt{3}}{(2\,\pi)^3} \int \frac{d\vec{k}_i}{e_{k_i}}\;
\Big(\frac{m_q}{e_{k_i}}\; \tilde{\phi}(k_i) \Big)\; \; 
\nonumber \\ &&\hspace*{-1cm}\times
\frac{\sqrt{3}}{(2\,\pi)^3} \int \frac{d\vec{k}_f}{e_{k_f}}\; 
\Big(\frac{m_q}{e_{k_f}}\; \tilde{\phi}(k_f) \Big)
\;\;\;\; \frac{4}{9}\; \frac{1 }{
\Big(1\!-\!\hat{Q}\cdot\frac{\vec{k}_i}{e_{k_i}}\Big)\;
\Big(1\!+\!\hat{Q}\cdot\frac{\vec{k}_f}{e_{k_f}}\Big)}.  
\label{twobod3}
\end{eqnarray}
We notice that this expression contains factors depending on the angle 
of $\vec{Q}$ and $\vec{k}$, which apparently have no counterpart
in the standard expectation, eq. (\ref{asymp}). 
 As this one was derived in a front-form approach, we will come back to it 
after considering the corresponding expression of $F^{(2)}_1(Q^2)$. 

\noindent
$\bullet$ {\it Expression of $F^{(2)}_1(Q^2)$ in the standard front form ($q^+=0$)}\\
The expression of the two-body contribution in the standard front-form 
approach  ($q^+=0$) is expressed in terms of the variables commonly used 
in this formalism, the Bjorken variables $x$ and the transverse momenta, 
$k_{\perp}$. It reads:
\begin{eqnarray}
F^{(2)}_1(Q^2)\!\!&=& \!\!  
\frac{1}{(2\,\pi)^3} \int \frac{d^2k_{i\perp}\;dx_i}{2\,x_i\,(1\!-\!x_i)}\;
  \Big(\frac{m_q}{e_{k_i}}\; \tilde{\phi}(k_i) \Big)\; \; 
\frac{1}{(2\,\pi)^3} \int \frac{d^2k_{f\perp}\;dx_f}{2\,x_f\,(1\!-\!x_f)}\; 
  \Big(\frac{m_q}{e_{k_f}}\; \tilde{\phi}(k_f) \Big)
\nonumber \\
&&\times \frac{4\pi\,\alpha_s\;\frac{4}{3}}{\mu^2+ \cdots}\;
\Big(I^{-}_{ps}+I^{-}_{pv}+I^{+}_{ps}+I^{+}_{pv}\Big)
+ \;({\rm  term:} \;i \leftrightarrow f) \, ,  
\label{twobod4}
\end{eqnarray}
where the quantities corresponding to the gluon propagator,  
and to the current, $I^{-}_{ps}+I^{-}_{pv}+I^{+}_{ps}+I^{+}_{pv}$, 
are given by:
\begin{eqnarray}
&& \mu^2+\cdots= \mu^2-(p_{i2}\!-\!p_{f2})^2= 
\nonumber \\
&&\hspace*{1cm}\mu^2+m_q^2\;\frac{(x_i\!-\!x_f)^2}{x_i \; x_f}+
\frac{\Big(x_i\;k_f \!-\! x_f\;k_i\Big)_{\perp}^2}{x_i \; x_f}
+2\,\Big(x_i\;k_f\!-\!x_f\;k_i\Big)_{\perp}
\ccdot Q_{\perp} + x_i\;x_f\;Q^2 \, ,
\nonumber \\
&&I^{-}_{ps}=I^{-}_{pv}=0\,,
\nonumber \\
&&I^{+}_{ps}+I^{+}_{pv}=
\frac{x_i\;\Big(x_i\;Q^2 \!-\! k_{i\perp} \ccdot Q_{\perp}\Big)}{2\,
\Big( m_q^2+\Big(k_i\!-\!x_i\;Q\Big)_{\perp}^2\Big)}\, .
\label{twobod5}
\end{eqnarray}
Two results stem from the above two equations. The contribution 
of the two-body current to the charge form factor vanishes at $Q^2=0$, 
as expected from the fact that two-body currents should not contribute 
to this quantity in a RQM approach. The expression of the asymptotic 
form factor can be obtained  by assuming that the wave function is peaked 
at low values of the $k$ variable, allowing one to discard $k$ terms 
in the matrix element of the current and the gluon propagator. 
One thus successively gets $I^{+}_{ps}+I^{+}_{pv}\simeq 1/2$, 
$ \mu^2+\cdots\simeq  x_i\;x_f\;Q^2 $ and: 
\begin{eqnarray}
F^{(2)}_1(Q^2)_{Q^2\rightarrow \infty}&=& 16\,\pi \;\frac{\alpha_s }{Q^2} \;
\frac{\sqrt{3}}{(2\,\pi)^3} \int \frac{d^2k_{i\perp}\;dx_i}{2\,x_i\;(1-x_i)}\;
  \Big(\frac{m_q}{e_{k_i}}\; \tilde{\phi}(k_i) \Big)\; \; 
  \nonumber \\  && \hspace*{1cm}\times
\frac{\sqrt{3}}{(2\,\pi)^3} \int \frac{d^2k_{f\perp}\;dx_f}{2\,x_f\;(1-x_f)}\; 
  \Big(\frac{m_q}{e_{k_f}}\; \tilde{\phi}(k_f) \Big)
\; \frac{1}{9}\; \frac{1 }{x_i\;x_f}.  
\label{twobod6}
\end{eqnarray}
\noindent
$\bullet$ {\it Comments}\\
The asymptotic expressions in the standard instant and front forms, 
eqs. (\ref{twobod3}) and (\ref{twobod6}), differ from each other. 
By making an appropriate change of variable however, it can be checked 
that they coincide. On the other hand, the presence of the factors 
$9/4\,(1\!-\!\hat{Q}\cdot\frac{\vec{k}_i}{e_{k_i}})^{-1}\;
(1\!+\!\hat{Q}\cdot\frac{\vec{k}_f}{e_{k_F}})^{-1}$ 
in eq. (\ref{twobod3}) and $(9\,x_i\;x_f)^{-1}$ in eq. (\ref{twobod6}) 
makes them apparently different from the usual  asymptotic expression, 
eq. (\ref{asymp}). Looking for an explanation of this possible discrepancy, 
it was found that the above expectation assumes that the integrand 
entering the expression of the pion-decay constant varies like $x\,(1-x)$. 
In such a case, which supposes that our wave function $\phi_0(k)$ 
exactly scales like $e_k^{-7/2}$, a full agreement is recovered. 
Actually, our expressions are more complete with some respects. 
For a part, the wave function contains some log corrections due 
to higher-order effects in the interaction. As for the extra factors 
mentioned above, they  were obtained in the past on a different basis 
(see for instance ref. \cite{Efrimov:1980}). 

The two-body contribution to the pion charge form factor assumes a quite
different expression in the standard instant- and front-form approaches.
Similarly to the one-body contribution, where we could find a common expression
depending on the orientation of the hyperplane which the physics is formulated
on, a common expression can be obtained for the two-body part. 
This one, which can be used for further investigations, is given in the 
appendix \ref{appc}. With this respect, it is noticed that the asymptotic
behavior is always produced by the pseudo-vector  pseudo-vector term of the
Fierz-transformed one-gluon exchange interaction, somewhat similarly 
to what has been shown for a long time in field-theory based approaches 
\cite{Maris:1998}, but, depending on the RQM approach, it
originates from that part involving the negative-energy component of the
intermediate particle with momentum $p^{*\,\mu}$ (instant form in the Breit
frame) or from the positive-energy one (front form with $q^+=0$). 
Interestingly, in an instant-form calculation away from the Breit frame but
preserving the condition $E_i=E_f$, or equivalently 
$\vec{Q} \perp (\vec{P}_i+\vec{P}_f)$, the part involving the negative-energy 
component tends to vanish with an increasing average momentum  
($|\vec{P}_i+\vec{P}_f| \rightarrow \infty $). The asymptotic contribution 
is then obtained from the other part with a positive-energy component. 
This result confirms the observation that 
a front-form calculation is closely related to an instant-form one 
in the above large momentum limit.
 
\noindent
$\bullet$ {\it Expression of the charge radius}\\
An expression of the squared charge radius is currently referred to 
in the literature: 
\begin{equation}
r^2_{\pi}=  \frac{3 }{4\,\pi^2\;f^2_{\pi}}=0.342\; {\rm fm}^2.
\label{msr}
\end{equation}
It has been derived in different ways
\cite{Tarrach:1979,Gerasimov:1979,Bernard:1988}.
In the original work by Tarrach for instance, it was obtained from 
the pion-quark-antiquark amplitude determined by the 
coupling constant, $g_{\pi qq}$. The relation to the above result assumes  
the equality $1/f_{\pi}=g_{\pi qq}/m_q$. Being sometimes presented 
as a consequence of the Goldstone-boson nature of the pion,
the question arises of whether this result could be recovered in a RQM approach.

Examining the derivation of the above approach in an instant-form 
approach (Breit frame), we found that a half is contributed by what 
corresponds to the Darwin-Foldy contribution to the form factor 
in the non-relativistic limit ($-Q^2/8\,m^2_q$), which involves 
positive-energy spinors, and the other half by a similar term 
involving negative-energy spinors accounted for by two-body currents. 
There are other contributions which respectively increase and decrease 
the above ones. They however cancel each other and have therefore 
no effect on the total result. These various results could thus be usefully 
compared to those obtained in RQM approaches by isolating the appropriate
contributions. A possible problem may concern the overall factor $1/f^2_{\pi}$ 
in eq. (\ref{msr}), which does not appear explicitly in these approaches. 
It has to be hoped that the description we are using for the pion 
is numerically close to that one determined by the coupling $g_{\pi qq}$.

\section{Results in the one-particle current approximation}
\begin{figure}[htb]
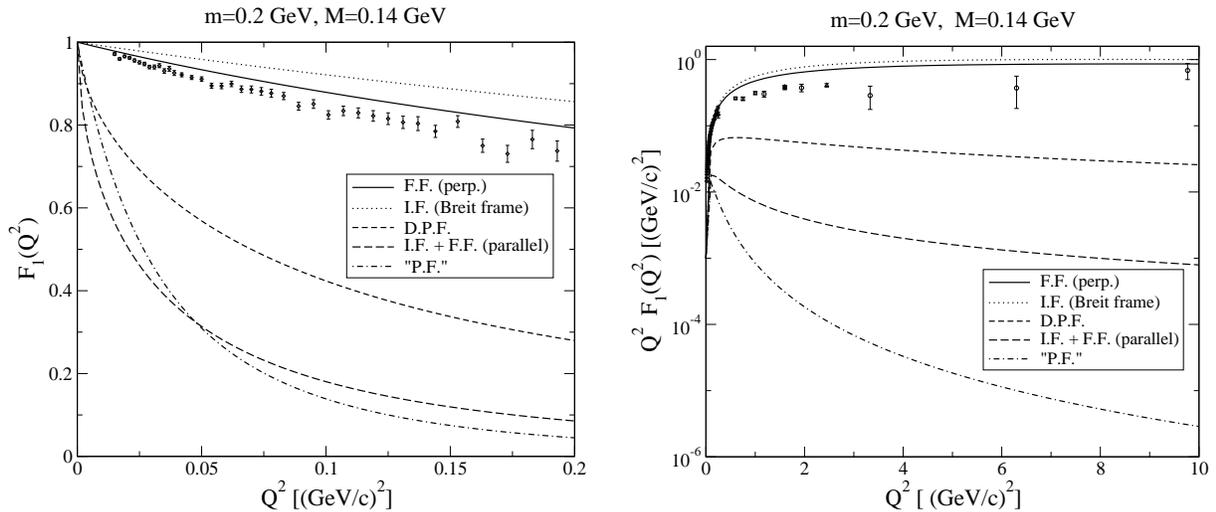
 
\epsfig{file=pi1s.eps, width = 7.7cm}  \hspace*{0.3cm}
\epsfig{file=pi1S.eps, width = 7.7cm} 
\caption{Pion charge form factor, $F_1(Q^2)$,  at small and high $Q^2$ 
(left and right panels respectively). In the latter case, the form factor is
multiplied by $Q^2$ and represented on a logarithmic scale 
to emphasize the asymptotic behavior. \label{F1}}
\end{figure} 

\begin{figure}[htb]
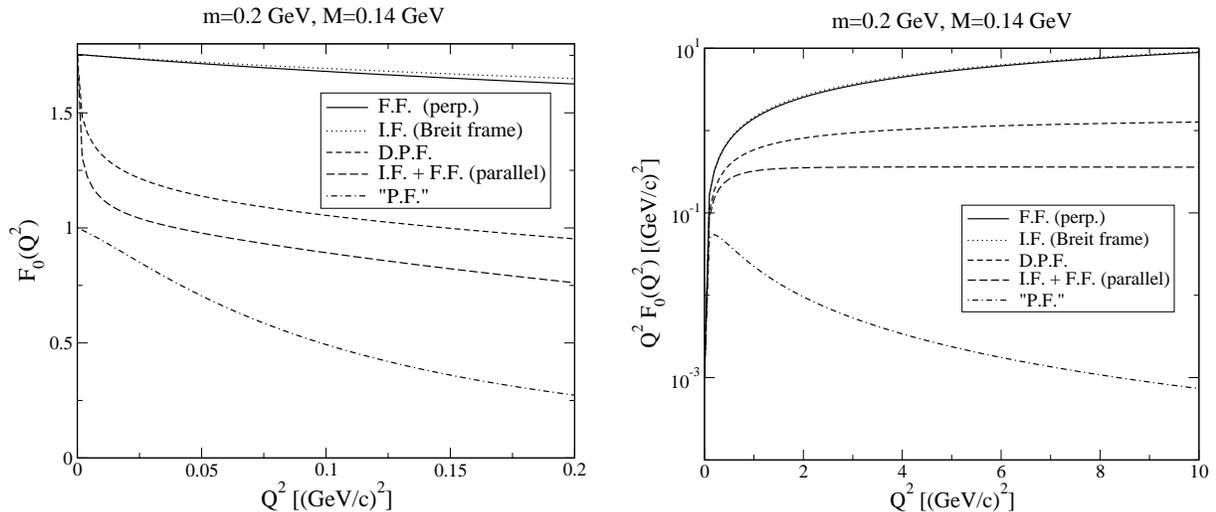
 
\epsfig{file=pi0s.eps, width = 7.7cm}   \hspace*{0.3cm}
\epsfig{file=pi0S.eps, width = 7.7cm} 
\caption{Same as for fig. \ref{F1} but for the scalar form factor $F_0(Q^2)$.
 \label{F0}}
\end{figure} 

We present in this section results for both the charge and scalar 
pion form factors calculated in the single-particle current approximation 
and for different RQM approaches. These ones include the instant 
and front forms with a standard momentum configuration, Breit frame in the former 
case (denoted I.F. (Breit frame)) and $q^+=0$ in the latter one 
(denoted F.F. (perp.)). Other frames are considered with 
a parallel momentum configuration (denoted I.F. + F.F. (parallel)).
Results also include an approach inspired from Dirac's point form 
(denoted D.P.F.) as well as an earlier ``point form" that has been employed 
in previous works (denoted ``P.F.").  
Contrary to the other results, these ones are frame independent. 
As mentioned in sect. 3, expressions of form factors can be obtained 
by generalyzing those given elsewhere for the scalar-constituent case 
\cite{Desplanques:2004sp,Desplanques:2008}.

Numerical results are shown altogether in figs.  \ref{F1} and \ref{F0} 
for the charge and scalar form factors respectively. 
Two ranges of momentum transfers are considered in each case,
low $Q^2$ on the l.h.s. ($0 - 0.2 \,({\rm GeV/c})^2)$ 
and high $Q^2$  on the r.h.s. ($0 - 10 \,({\rm GeV/c})^2$). 
The first one is expected to evidence a sensitivity to the charge 
(or scalar) radius while the second one {\it a priori} emphasizes 
the asymptotic behavior. As this one should behave like $Q^{-2}$, 
the corresponding form factors are multiplied by $Q^2$  so that 
the product be close to a constant in the limit of large momentum 
transfers (up to log terms). Measurements for the
charge form factor are also shown in fig. \ref{F1}.

The comparison of different theoretical approaches evidences features quite
similar to the scalar-constituent case  \cite{Desplanques:2004sp}. The Breit-frame
instant-form results slightly overshoot the front-form ones with $q^+=0$. 
The other instant- and front-form results with a parallel momentum configuration, 
which coincide with each other, show a fast drop off at small $Q^2$, 
a property that is also found for the point-form results. 
As explained elsewhere \cite{Amghar:2003tx}, this drop off points 
to a squared charge radius that is determined by the inverse 
of the squared pion mass and is larger than experiment by an order 
of magnitude. For a part, this sensitivity to the pion mass explains 
the large discrepancies between results shown at the r.h.s. of figs. 
 \ref{F1} and \ref{F0}, which roughly scale like in the scalar-constituent case. 

The comparison with experiment, which is only possible for the charge form
factor, shows that the standard instant- and front-form approaches provide
reasonable results as a starting point while all the other ones are off,
sometimes by orders of magnitude. This is again somewhat similar 
to the scalar-constituent case where the role of the experiment is played 
by an exact calculation. As has been suggested \cite{Desplanques:2004sp} and
shown later on \cite{Desplanques:2008}, 
the largest discrepancies point to a violation of Poincar\'e space-time 
translation invariance, which requires specific two-body currents 
to be restored. This does not necessary imply that the other results 
in the standard instant- and front-form approaches are fully under control.

Looking carefully at the corresponding product of the charge form factor
with $Q^2$, shown at the r.h.s. of fig. \ref{F1} , it is found that 
the appearance of a plateau, which could suggest that the expected 
asymptotic behavior  is reached, is misleading. The comparison 
with the scalar form factor, shown at the r.h.s. of fig. \ref{F0},  
is here very instructive. In this case, the same product increases, 
in agreement with possible log$(Q^2)$ deviations and, moreover, 
is significantly larger. Globally, it thus appears 
that the above charge form factors at high $Q^2$ are suppressed
with respect to the scalar ones, which can be better seen by examining
results for much higher momentum transfers (see next section). 
The origin of the suppression, roughly given by a factor $Q^2$, 
resides for a part in the coupling of quarks to the external probe, 
as can be checked from considering matrix elements of the current 
between positive-energy spinors \cite{Amghar:2003tx}. 
A similar suppression,  but within a truncated field-theory light-front 
calculation, was mentioned in ref. \cite{Carbonell:1998}. This result 
represents an important difference with the scalar-particle case. 

At this point, a few further remarks can be made. The spin-1/2 nature 
of the constituents also shows up in results at small $Q^2$. 
Its effect in the better cases (standard instant- and front-form 
approaches) is found to be very similar to the one produced 
by the Melosh transformation in other works \cite{Chung:1988mu}, 
with an increased contribution to the charge radius (Darwin-Foldy
term). Results evidence some sensitivity to the dynamics like the
description of the confinement. This has not been considered 
in detail here but we notice that reproducing the pion-decay constant 
allows one to reduce the uncertainty.

\section{Results with two-body currents and asymptotic behavior}

\begin{figure}[htb]
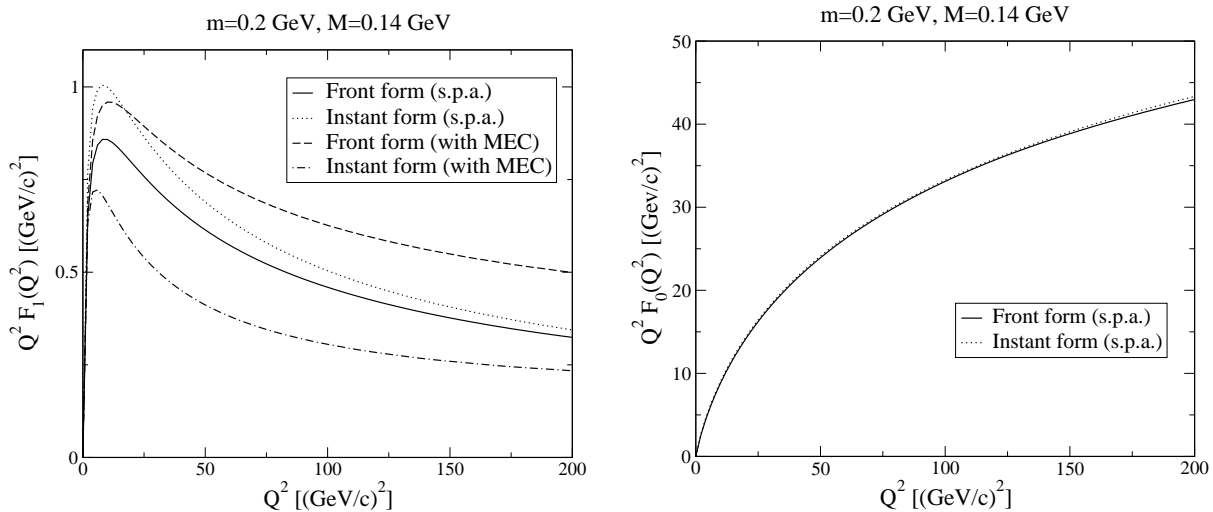
 
\epsfig{file=pi1SS.eps, width = 7.7cm}  \hspace*{0.3cm}
\epsfig{file=pi0SS.eps, width = 7.7cm} 
\caption{Form factors $F_1(Q^2)$ and $F_0(Q^2)$ in the ultra asymptotic 
domain: contribution from the single-particle current (s.p.a.) 
and, for  $F_1(Q^2)$, contribution incorporating two-body currents (with MEC).
Calculations are performed in the standard instant and front-form 
approaches.
\label{pair1}}
\end{figure} 
\begin{figure}[htb]
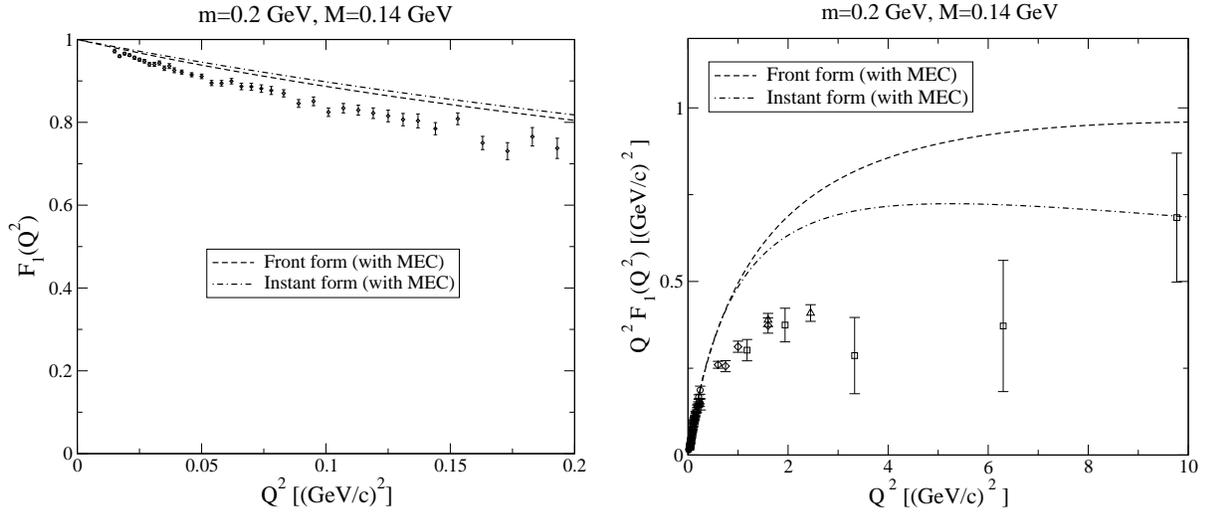
 
\epsfig{file=pipa1s.eps, width = 7.7cm}\hspace*{0.3cm}
\epsfig{file=pipa1S.eps, width = 7.7cm} 
\caption{Effect of two-body currents on the charge form factor, $F_1(Q^2)$, 
at low and intermediate momentum transfers: results are presented 
for the standard instant and front-form approaches. 
\label{pair2}}
\end{figure} 
We consider in this section  the pion charge form factor with including the
contribution of two-body currents. This is done for both the instant-form and
front-form approaches, respectively in the standard Breit-frame and 
so-called $q^+=0$ configurations. As already mentioned, the other approaches 
suppose dominant contributions from specific two-body currents restoring 
the Poincar\'e  space-time translation invariance. These ones could 
be determined along lines developed in ref. \cite{Desplanques:2008} but it is
not clear how they will affect the contribution of two-body currents of interest
here (possible double counting). 

As two-body currents considered in this section are mainly motivated 
by the asymptotic behavior of the charge form factor, we first present 
results relevant to this feature. These ones, multiplied by $Q^2$, 
are given in fig. \ref{pair1} (l.h.s.) for a range of $Q^2$ extending 
from 0 to 200 (GeV/c)$^2$. For the
sake of the discussion, we also present in this figure (r.h.s.) results
pertinent to the scalar form factor. The two-body currents also contribute
at low and intermediate values of $Q^2$. The corresponding results 
for the charge form factor alone are shown in fig.  \ref{pair2}. 
The range under consideration is the same as in fig. \ref{F1}, but 
a linear scale instead of a logarithmic one is adopted for the r.h.s. part.

We first notice that, contrarily to what  the consideration 
of the r.h.s.  of fig. \ref{F1} would suggest, the asymptotic behavior 
of the charge form factor is not reached yet. 
Examination of fig. \ref{pair1} (l.h.s.) clearly shows that 
the product  of $Q^2$ with the charge form factor calculated 
from the single-particle current, after evidencing a maximum 
around  $10$ (GeV/c)$^2$, begins to slowly decrease. 
The difference with the scalar form factor shown in the r.h.s. 
of the same figure, which has the right asymptotic behavior, is striking. 
Accounting for the contribution of two-body currents to the charge form factor 
allows one to obtain its asymptotic behavior which is expected to be given
numerically by $F_1(Q^2)_{Q^2\rightarrow \infty}=$ 0.17 (GeV/c)$^2$$/Q^2$ 
for our pion description. As it can be seen on fig. \ref{pair1} 
around  $200$ (GeV/c)$^2$, this value is not far to be reached in the 
instant-form case but still rather far away in the front-form one. 
The different behavior can be explained as follows. 
The instant-form two-body current implies a contribution originating 
from the pseudo-scalar part of the interaction ($I^{-}_{ps}$), 
which  cancels the single-particle one, leaving as a neat result 
the contribution due to the pseudo-axial vector part ($I^{-}_{pv}$), 
while this destructive contribution is absent in the front-form case. 
A few further comments may provide better insight on the above results. 
First, in spite of different formalisms, present results qualitatively agree 
with those of earlier works \cite{Maris:1998,Braguta:2008}. Some of the quantitative
differences, especially the onset of the asymptotic behavior at larger $Q^2$ 
in this work, can be related for a part to our pion description 
that involves sizeable high-momentum components. 
Second, the difference between the asymptotic value of the
form factor given above in the text and the current one given 
by eq. (\ref{asymp}), 0.11 (GeV/c)$^2$$/Q^2$,  can be ascribed 
to the non-perturbative effects mentioned in the comments 
about this last result (end of sect. 3.3). 

Considering now the results for the charge form factor at very small 
transfers (l.h.s. of fig.  \ref{pair2}), it is observed that the contribution 
of the two-body currents tends to remove the difference between the instant- 
and front-form form factors. Such a result is not totally unexpected as, for a
part, the two-body currents tend to restore Lorentz invariance in first place
and perhaps some Poincar\'e space-time translation invariance in second place. It actually
holds for the contribution of the pseudo-scalar part of the interaction alone, 
$I^{-}_{ps}$ in eq. (\ref{twobod2}), what it should as the part related to the
pseudo-axial vector part, $I^{-}_{pv}$, is irrelevant for the argument. 
It remains that some contribution could be due to the confinement 
interaction but its non-perturbative character does not allow one 
to determine how much it affects  $I^{-}_{ps}$ in the instant form. 

To some extent, the above Lorentz-invariance argument applies to the 
form factor in the intermediate range, shown in the  r.h.s. of 
fig.  \ref{pair2}, but the effect is larger than needed. We observe 
that the contribution of two-body currents tends to make the instant-form 
form factor closer to experiment while it makes the front-form one 
further away. 
To better understand the role of two-body currents depending 
on the form, it is useful to consider separately the two contributions 
in the instant form, $I^{-}_{ps}$ and $I^{-}_{pv}$, eq. (\ref{twobod2}).  
The first one, which has no well determined counterpart in the 
front-form case, is negative over the full range of $Q^2$ considered 
in this work and decreases faster than $Q^{-2}$. The second one,
which is positive and provides the $Q^{-2}$ asymptotic behavior, 
is close to the contribution $I^{+}_{ps}+I^{+}_{pv}$ calculated 
in the front-form approach. 

Having considered two-body currents, we can now discuss the pion charge
radius to which they can contribute. This is done in relation
with the expression  $r^2_{\pi}=  3 /(4\,\pi^2\;f^2_{\pi})$, 
which gives the value $r^2_{\pi}=0.342\; {\rm fm}^2$ for  
$f_{\pi}=93$ MeV. A value that could be more relevant for the discussion 
is $r^2_{\pi}=0.263\; {\rm fm}^2$, which corresponds to 
$f_{\pi}=106$ MeV obtained with our pion description. 
In the instant form, it is expected that part of the contribution should 
have a counterpart in the piece of the two-body currents associated 
to the pseudo-scalar component, $I^{-}_{ps}$. We first notice that 
the squared matter radius has a relatively small role, of the order 
of $0.08\,{\rm fm}^2$ in the present calculation to be compared 
to the measured value of the squared charge radius, $0.43\,{\rm fm}^2$. 
We also notice that the part of two-body currents due to the pseudo-axial 
vector component of the interaction, also small ($-0.02\,{\rm fm}^2$),
is irrelevant for a comparison with the above prediction which assumes 
a pseudo-scalar component and a point-like pion. As already mentioned, this approximate 
prediction should be compared to that part in our calculations 
which corresponds to the Darwin-Foldy term. Isolating this term 
in the instant-form results, we find $0.15\,{\rm fm}^2$ 
for the one-body current and $0.08\,{\rm fm}^2$ for the two-body one. 
The sum, $0.23\,{\rm fm}^2$, compares to the expectation $0.263\,{\rm fm}^2$.  
In the front form, where we only have a contribution from the one-body 
current, a similar procedure gives $0.29\,{\rm fm}^2$, 
which also compares to the expectation $0.263\,{\rm fm}^2$. Thus, there is 
a reasonable agreement  of the present results for the charge radius 
and the approximate expectation advocated in many works. 
Another relation, suggested by the analysis of various contributions 
to the above expression of $r^2_{\pi}$, is worthwhile to be mentioned. 
The usual Darwin-Foldy correction to the form factor in the instant form 
takes the form $-Q^2/(8\,m_q^2)$ (non-relativistic limit). 
When the similar term associated here to the two-body current is
considered, one gets instead $-Q^2/(4\,m_q^2)$. This last term can 
be seen as the first term in the $Q^2$ expansion of the vector-meson 
dominance model of form factors,  $-Q^2/m_V^2$, where 
$m_V^2= 4\,m_q^2+\cdots$. 

When considering calculations of the two-body current contribution, 
a large sensitivity to various ingredients was found. Most of it 
points to the role of higher-order effects in the interaction which, 
for consistency, were neglected. These effects nevertheless 
show up in looking at some of the results presented here. As an 
example, it is likely that  the charge and the scalar form factors 
in the asymptotic regime should be comparable, up to a factor 1-2.
Examination of results presented in fig.  \ref{pair2} rather suggest 
that they differ by a factor 10-20. This can be largely explained by 
non-perturbative corrections to the wave function entering 
the calculation. Thus, these effects lead to log($Q^2/m_q^2$) terms 
that enhance the scalar form factor in the asymptotic domain while 
they are ignored in the present derivation of two-body currents based 
on a perturbative approach. An other hint about these non-perturbative
effects is provided by the comparison of the contribution 
of two-body currents in different forms. While the contributions 
of the single-particle current in the standard instant- and front-form 
approaches are relatively  close to each other, those for the two-body 
currents significantly differ. Most of the discrepancy can be traced back 
to the term  $I^{-}_{ps}$, whose role exceeds the restoration of Lorentz
invariance at soon as the momentum transfer increases.

\section{Discussion and conclusion}
We first considered in this work the single-particle contribution 
to the charge and pion form factors in different forms of RQM approaches.
These ones include the standard instant and front forms (Breit frame and 
the momentum configuration $q^+=0$ respectively), the same forms with a ``parallel" momentum configuration and
the point form.  The interaction entering the mass operator is chosen as the
sum of a confining interaction and a gluon-exchange one, which is {\it a
priori} essential for getting the right asymptotic behavior of form factors. 
For the case of the charge form factor, where measurements are available, 
the standard instant- and front-form approaches provide results relatively
close to them. All the other results fall far apart, as a result of their
dependence on the momentum transfer through the ratio $v=Q/(2\,M_{\pi})$. 
Thus the pattern evidenced by these results is very similar to that one 
obtained with a model of scalar particles \cite{Desplanques:2004sp}. 
As has been shown, the  last results evidence a strong violation 
of Poincar\'e space-time translation invariance while this property 
is approximately fulfilled by the standard instant- and front-form 
ones. This discards the underlying approaches as efficient ones 
to describe the pion properties.

While there is some similarity of the above results with earlier ones 
for a theoretical model with scalar constituents, there are nevertheless 
differences. The most striking one 
concerns the ratio of the charge and scalar form factors. Whatever 
the approach, the first one is suppressed compared to the other one 
at high $Q^2$, preventing one from reproducing its expected asymptotic 
behavior, $Q^{-2}$. This result, which seems to characterize relativistic 
quantum mechanics approaches with a single-particle current, implies the 
spin-1/2 structure of the quarks. This conclusion is 
comforted by the consideration of the scalar form factor which,  
up to a finite coefficient, has the right power-law behavior.

We then looked at the two-body currents whose contributions 
are necessarily required to reproduce the asymptotic behavior 
of the charge pion form factor expected from QCD. This has been done 
in both the standard instant and front forms, which are the only 
approaches to provide a reasonable starting point for further 
improvement. The two-body currents have been derived on the basis 
that adding their contribution  to that one gene\-rated by the 
wave function should allow one to recover the full Born amplitude. 
While the description of hadron properties at low- and high-momentum 
transfers are often con\-si\-dered as disconnected, we showed that 
the above two-body currents were able to  reproduce the expected 
expression of the pion charge form factor in the asymptotic regime. 
This result involves the pseudo-vector  pseudo-vector  part 
of the Fierz-transformed one-gluon exchange interaction, in agreement with 
what is expected in field-theory based approaches \cite{Maris:1998}. 
Quantitatively, the relevance of this result appears at quite 
high momentum transfers, pro\-ba\-bly beyond the range 
where the pion charge form factor has been measured. At very 
low-momentum transfers, the contribution of these two-body currents 
depends on the form. 
While it increases the front-form form factor, as at high $Q^2$, 
it decreases the instant-form one. The total effect compares 
to the difference of the instant- and front-form results 
in the single-particle current approximation, which it tends to cancel. 
Such a result is expected for a part as far as the two-body currents 
we are considering should contribute to restore Lorentz invariance. 
Between the very low- and  very high-$Q^2$ domain, it is difficult 
to attribute a specific role to the above two-body currents, 
but it does not seem to affect much the comparison with experiment. 

We have not made any attempt to provide a better description 
of the measured pion charge form factor, being mainly concerned 
by the determination of an approach where the bulk contribution 
would be given by a one-body current so that the derivation 
of two-body currents of interest here be simplified as much as possible. 
Following refs. \cite{Cardarelli:1994,Cardarelli:1995dc}, the  introduction 
of some quark form factor could reduce the discrepancy with experiment. 
One can nevertheless guess that a smaller quark mass would contribute to get
both a better value of $f_{\pi}$ and a better charge radius, 
hence a better form factor at very small  $Q^2$. 
A pion description with a smaller amount of large momentum components could
also have a similar effect. As, apart from the quark mass, the parameters
entering the mass operator we used are essentially fixed, 
the only issue would be to modify the form of this operator. 
One can expect for instance that retardation effects, ignored here, 
could contribute to reduce the strength of the one-gluon exchange 
at short distances \cite{Amghar:2000,Amghar:2000pc} and, consequently, 
the interaction at high momenta. Thus, 
while  relativistic quantum mechanics is not the most fundamental approach 
for describing the pion properties, getting these ones relatively well 
in this framework does not seem to be out of reach.

\noindent
{\bf Acknowledgement}\\
We are very grateful to O. Leitner for useful comments about recovering the
asymptotic behavior of the pion charge form factor in the present approach. 

\appendix
\section{Accounting for the spin of the constituents in deriving 
the mass operator}
\label{appa}
The simplest expression of the RQM pion wave function reads;
\begin{equation}
\psi(p_1,p_2)\propto \bar{u}(p_1)\gamma_5 v(p_2)\; \frac{m_q}{e_k} \;\phi_0(k),
\label{appa:1}
\end{equation}
where the variable $k$ is related to the constituent momenta by the relation 
$4\, e_k^2=(p_1+p_2)^2$. The spinors are normalized as follows:
\begin{equation}
\sum_{spins} u(p)\bar{u}(p)=\frac{\gamma \cdot p +m_q}{2m_q}, \;\;\;
\sum_{spins} v(p)\bar{v}(p)=\frac{\gamma \cdot p -m_q}{2m_q}.
\label{appa:2}
\end{equation}
Other expressions of the pion wave function reduce to the above one, 
eq. (\ref{appa:1}), taking into account that the on-mass shell character 
of particles in the RQM framework allows one to use the Dirac equation 
for free particles. In principle, $\psi(p_1,p_2)$ is a solution 
of a wave equation which should be cast into the form of a mass operator 
equation by making some change of variable. The problem and the related
constraints were considered for scalar particles, see ref. \cite{Amghar:2002jx} 
for an example. When considering a standard meson-exchange 
interaction like the gluon one in the present case, the non-zero spin of
constituents has to be accounted for, raising further problems. 
To illustrate them, it is instructive to look at the action 
of the spin part of this interaction together with that one 
for the pion wave function: 
\begin{equation}
V\;\psi(p'_1,p'_2) \propto \sum_{spins} \bar{u}(p_1)\gamma^{\mu} u(p'_1)\;
\bar{v}(p'_2)\gamma_{\mu} v(p_2)
\;\bar{u}(p'_1)\gamma_5 v(p'_2).
\label{appa:3}
\end{equation}
Using a Fierz transformation, the spin part of the interaction 
can also be written:
\begin{eqnarray}
\bar{u}(p_1) \gamma^{\mu} u(p'_1)\; \bar{v}(p'_2) \gamma_{\mu} v(p_2)=
\bar{u}(p_1) v(p_2)\; \bar{v}(p'_2)u(p'_1)
+\bar{u}(p_1) i\gamma_5 v(p_2)\; \bar{v}(p'_2) i\gamma_5 u(p'_1)
\nonumber \\
-\frac{1}{2}\bar{u} (p_1)\gamma^{\mu} v(p_2)\; 
\bar{v}(p'_2) \gamma_{\mu}  u(p'_1)
-\frac{1}{2}\bar{u}(p_1) \gamma^{\mu}\gamma_5 v(p_2)\; 
\bar{v}(p'_2) \gamma_{\mu}\gamma_5  u(p'_1),
\label{appa:4}
\end{eqnarray}
where one recognizes a term with the same structure as the input 
wave function ($\bar{u}\gamma_5 v$) and another one with a pseudo-vector 
character ($\bar{u}\gamma_{\mu}\gamma_5 v$). When performing the summation 
over spins in eq. (\ref{appa:3}), it is found that these two terms contribute, 
with the result:
\begin{eqnarray}
&&\sum_{spins} \bar{u}(p_1)\gamma^{\mu} u(p'_1)\;
\bar{v}(p'_2)\gamma_{\mu} v(p_2)
\;\bar{u}(p'_1)\gamma_5 v(p'_2)=
\nonumber \\  && \hspace*{2cm}
\bar{u}(p_1) \gamma_5 v(p_2)\; \frac{(p'_1 \cdot p'_2 + m_q^2)}{m_q^2}
-\frac{1}{2}\;  \bar{u}(p_1) \gamma^{\mu}\gamma_5 v(p_2)\;
\frac{(p'_1+p'_2)_{\mu}}{m_q}.
\label{appa:4bis}
\end{eqnarray}
The first term at the r.h.s., where the dependence on variables relative 
to the initial and final states factorizes, does not provide any difficulty 
in reducing a wave equation to a mass operator. The second one does however. 
Writing $(p'_1+p'_2)_{\mu}=(p'_1+p'_2-p_1-p_2)_{\mu}+(p_1+p_2)_{\mu}$ 
and using the Dirac equation, the above equation also reads:
\begin{eqnarray}
&&\sum_{spins} \bar{u}(p_1)\gamma^{\mu} u(p'_1)\;
\bar{v}(p'_2)\gamma_{\mu} v(p_2)
\;\bar{u}(p'_1)\gamma_5 v(p'_2)=
\nonumber \\  && \hspace*{2cm}
\bar{u}(p_1) \gamma_5 v(p_2)\; \frac{p'_1 \cdot p'_2}{m_q^2}
-  \bar{u}(p_1) \gamma^{\mu}\gamma_5 v(p_2)\;
\frac{(p'_1+p'_2-p_1-p_2)_{\mu}}{2\,m_q},
\label{appa:4ter}
\end{eqnarray}
where the last term is seen to  depend on the surface on which physics is
described (through $\xi^{\mu}$), 
quite in the same way as a meson propagator would generally. 
The problem is well known in relativistic quantum mechanics. It points 
to the constraints that the interaction entering the mass operator 
has to fulfill. These ones implicitly account for higher-order 
corrections in the interaction, which cancel the above undesirable 
contributions. This is made possible by the fact that these contributions 
have an off-energy shell character (see eq. (\ref{boost2})). 
Discarding these ones, the relevant part of the gluon 
exchange appropriate for describing the pion in the RQM framework
may thus be written in the following factorized form:
\begin{equation}
V\; \propto g^2 \frac{\bar{u}(p_1)i\gamma_5 u(p_2)\;
\bar{v}(p'_2)i\gamma_5 v(p'_1)}{(\vec{k}-\vec{k}')^2}\;
\Big(2- \frac{m^2_q\,(e^2_k+e^2_{k'})}{2\,e^2_k\;e^2_{k'}}\Big)\, .
\label{appa:5}
\end{equation}
The last factor has been introduced to represent as much as possible the effect 
of the factor $p'_1 \cdot p'_2 \;(=e^2_{k'}+k'^2)$ in eq. (\ref{appa:4ter}),
which originates from the original vector-vector nature of the interaction
($\gamma^{\mu} \;\gamma_{\mu}$), while keeping the symmetry between $k$ and $k'$. 
It differs from the factor $(2-m_q^2/e^2_k)^{1/2}(2-m_q^2/e^2_{k'})^{1/2}$ 
introduced by Godfrey and Isgur \cite{Godfrey:1985} by off-energy shell 
corrections proportional to $m^2_q$, which are in any case 
part of the uncertainty  on the derivation of the mass operator. 
Due to the presence of square root factors however, the last choice 
may introduce further complication in the derivation of two-body currents. 
Whatever the choice, we notice that a pseudo-scalar pseudo-scalar  type 
interaction ($\gamma_5 \;\gamma_5 $) reproducing the non-relativistic limit 
of the one-gluon exchange would  provide a different result 
(a factor $e^2_{k'}$ instead of $e^2_{k'}+k'^2$ in the present case). 
Thus, the choice made here is consistent 
with the symmetry character of the interaction $V$, but could miss 
some off-shell corrections. Gathering all the relevant factors, 
the wave equation reads:
\begin{eqnarray}
&&(M^2-4\,e_k^2) \; \bar{u}(p_1)\gamma_5 v(p_2)\; \frac{m_q}{e_k} \;\phi_0(k)
= -\int \frac{d\vec{k}'}{(2\,\pi)^3}\;
 \frac{4 \;g^2\;\frac{4}{3}}{(\vec{k}-\vec{k}')^2} \;
\Big(2- \frac{m^2_q\,(e^2_k+e^2_{k'})}{2\,e^2_k\;e^2_{k'}}\Big) 
\nonumber\\  && \hspace*{2cm} \times  \frac{m_q^2}{\sqrt{e_k}\sqrt{e_{k'}}} 
\bar{u}(p_1)i\gamma_5 v(p_2)\;\;\frac{1}{2}\;\sum_{spins}
\bar{v}(p'_2)i\gamma_5 u(p'_1)\;
\;\bar{u}(p'_1)\gamma_5 v(p'_2)\; \frac{m_q}{e_{k'}}  \;\phi_0(k')
\nonumber\\ && \hspace*{0.3cm} 
=-\int \frac{d\vec{k}'}{(2\,\pi)^3}\;
 \frac{4 \;g^2\;\frac{4}{3}}{(\vec{k}-\vec{k}')^2} \;
 \frac{m_q^2}{\sqrt{e_k}\sqrt{e_{k'}}}\;
\bar{u}(p_1)\gamma_5 v(p_2) \; 
\Big(2- \frac{m^2_q\,(e^2_k+e^2_{k'})}{2\,e^2_k\;e^2_{k'}}\Big)
\frac{ e_{k'}^2}{m_q^2} 
\;\frac{m_q}{e_{k'}}  \;\phi_0(k'),
\nonumber\\
\label{appa:6}
\end{eqnarray}
which, after some simplification, is found to give eq. (\ref{x2a}) (apart from 
the confinement part not considered here). 
\section{High momentum behavior of solutions}
\label{appb}
The behavior of wave functions at high momenta, which in principle is relevant
for calculating form factors in the asymptotic regime, is determined by the most
singular part of the interaction. Examination of eq. (\ref{x2a}) indicates that
the problem relevant to the determination of this behavior amounts to look 
for solutions of the following equation:
\begin{equation}
(p-\frac{\alpha}{r})\; \tilde{\psi}(r)=0\, ,
\end{equation}
where $\tilde{\psi}(r)$ is the Fourier transform of the momentum 
wave function $\tilde{\phi}(k)=\sqrt{e_k}\;\phi_0(k)$ 
and $\alpha= 8\,\alpha_s/3$. Though the problem was not always emphasized, 
it is part of earlier works \cite{Carlson:1983rw,Godfrey:1985,Cardarelli:1994}.
Solutions can be obtained from the following relation:
\begin{equation}
p (r^{\mu-1})=\frac{\mu\;cotg(\frac{\mu\;\pi}{2})}{r} \;(r^{\mu-1})\, ,
\end{equation}
from which we infer that solutions in momentum space are given by:
\begin{equation}
\tilde{\psi}(p)\propto p^{-\mu-2}\, ,
\end{equation}
with $\alpha=\mu\;cotg(\frac{\mu\;\pi}{2})$. Particular cases are the following
ones:
\begin{eqnarray}
&&\alpha=\epsilon ,\;\mu= 1-\frac{2\,\epsilon}{\pi}:\;
\;\;(p-\frac{\epsilon}{r})\;
 r^{\frac{2\,\epsilon}{\pi}}=0, \;\;
 \tilde{\psi}(p)\propto p^{-3+\frac{2\,\epsilon}{\pi}}\, ,
\nonumber \\
&&\alpha=\frac{1}{2} ,\;\mu= 1/2:\;\;\;
(p-\frac{1}{2\,r})\; r^{-1/2}=0, \;\;
\tilde{\psi}(p)\propto p^{-2.5}\, ,
\nonumber \\
&&\alpha=\frac{2}{\pi} ,\; \mu= 0:\;\;\;
(p-\frac{2}{\pi\,r})\; r^{-1}=0, \;\;
\tilde{\psi}(p)\propto p^{-2},
({\rm limit \;case})\, .
\end{eqnarray}
The first case corresponds to a perturbative one which, in our case, implies 
the behavior $\tilde{\phi}(k)\propto k^{-3},\;\phi_0(k)\propto k^{-7/2}$. 
When the strength of the interaction increases, non-perturbative effects 
generally given by log terms can be large enough to change the asymptotic 
power, by half a unit for $\alpha=\frac{1}{2} $ and one unit for 
$\alpha=\frac{2}{\pi}$. As has been shown \cite{Leyouanc:1997},  the last 
case corresponds to the occurrence of a critical regime in the QED case. 
The relevance of this result for QCD and the spontaneous breaking 
of the chiral symmetry is not quite clear but we notice that the range 
for the strong coupling, $\alpha_s$, usually referred to is of the order 
or even exceeds the critical value of $\frac{2}{\pi}\simeq 0.64 $ 
($\alpha=1.33$ for $\alpha_s=0.5$). Possible problems related to the onset 
of a critical regime, that are generally ignored, appear here due 
to the quadratic form of the mass equation and the extra factor 
$\Big(2- (m^2_q\,(e^2_k+e^2_{k'}))/(2\,e^2_k\;e^2_{k'})\Big)$, which separately
enhance the strength of the force by a factor 2 at large momentum. 
They could be alleviated by the decreasing of $\alpha_s$ with the momentum 
transfer, or the effect of retardation effects. In practice, we will choose 
a value of $\alpha_s$ that corresponds to $\alpha=0.5$ at most 
({\it i.e.} $\alpha_s=3/16$).

\section{Two-body currents and asymptotic behavior}
\label{appc}
We consider in this appendix the derivation of two-body currents that are related 
to the one-gluon exchange and could contribute to the pion charge 
form factor in the asymptotic domain, of special interest in this work. 
This is done for the general case where physics is formulated 
on an arbitrary hyperplane of orientation $\xi^{\mu}$ and, for 
the following discussion, we introduce  ``positive" and ``negative" energies 
as defined with respect to this orientation. Particular applications 
concern both the instant-form and the front-form approaches.  
Essentially, the derivation amounts to include contributions that
are not accounted for in the single-particle contribution. This involves
contributions with intermediate particles carrying a ``negative" energy, which
are not part of a RQM formalism. Contributions with a ``positive" energy 
also occur but, in this case, they involve off-energy shell effects that the
formalism ignores. These ones are not necessarily independent of the previous
ones however. What corresponds to ``negative"-energy components 
in a given approach often appears as an off-energy shell in another one. 

Our starting point is the expression of the charge form factor given 
by eq. (\ref{y3a}). In order to derive the two-body currents we look for, 
the relation with the Born amplitude shown in fig. \ref{fig:2} has to be 
established. This is done by demanding that they coincide for the contribution
with a  ``positive" energy intermediate quark when higher orders in the
interaction are discarded. We notice that the above procedure fixes 
the strength of the Born amplitude entering the calculation providing 
a check on the ability of the underlying formalism to account for it. 
The contribution to the form factor thus obtained reads:
\begin{eqnarray}
\Delta F_1(Q^2)= 
\int \frac{d\vec{p}_{i2}}{(2\,\pi)^3}
\;\int \frac{d\vec{p}_{f2}}{(2\,\pi)^3}\;
\; \frac{\xi \ccdot (p_{i1}\!+\!p_{i2})}{2\,e_{i2}\;\;\xi \ccdot p_{i1}}
\;\Big(\frac{m_q}{e_{k_i}}\; \tilde{\phi}(k_i) \Big) 
\;\frac{ \xi \ccdot I}{ \xi \ccdot (p_{i1}\!+\!p_{i2}\!+\!p_{f1}\!+\!p_{f2})}
 \hspace{0.5cm}
\nonumber \\ \times \; 
\frac{1}{\Big(m_q^2\!-\!p_f^{*2} \Big)}\;
\frac{g^2\;\frac{4}{3}}{\mu^2\!-\!(p_{i2}\!-\!p_{f2})^2}\;
\;\frac{\xi \ccdot (p_{f1}\!+\!p_{f2})}{2\,e_{f2}\;\;\xi \ccdot p_{f1}}
\;\Big(\frac{m_q}{e_{k_f}}\; \tilde{\phi}(k_f) \Big)+ 
\;({\rm term:} \;i \leftrightarrow f )\, , 
\label{appc1}
\end{eqnarray}
where $p^{*\,\mu}_f=P_f^{\mu}-p_{i2}^{\mu}$, while $I^{\mu}$ is given by:
\begin{eqnarray}
I^{\mu}&=& \frac{1}{2}\; 8\,m_q^3\;
\nonumber \\ && \hspace*{-8mm}\times \,
 {\rm Tr} \Bigg[  \frac{m_q\!-\!\gamma \ccdot p_{i2}}{2\,m}\,i\gamma_5\,
 \frac{m_q\!+\!\gamma \ccdot p_{i1}}{2\,m_q} \; \gamma^{\mu}\;
 \frac{m_q\!+\!\gamma \ccdot p^{*}_f}{2\,m_q}\;\gamma^{\nu}\;
  \frac{m_q\!+\!\gamma \ccdot p_{f1}}{2\,m_q}\,i\gamma_5\,
   \frac{m_q\!-\!\gamma \ccdot p_{f2}}{2\,m_q} \;\gamma_{\nu}\;  \Bigg]
   \nonumber \\ 
&& \hspace*{-8mm}  = 2\,\frac{(m_q^2\!+\!  p_{f1}\!\cdot\! p_{f2})}{m_q^2}\;
 \Bigg[\!  p_{i1}\!\cdot\! p_{i2}\;p^{*\,\mu}_f
 - p_{i1}\!\cdot\! p^{*}_f\;p_{i2}^{\mu} +p_{i2}\!\cdot\! p^{*}_f\;p_{i1}^{\mu} 
 +m_q^2\;(p^{*\,\mu}_f\! +\! p_{i1}^{\mu} \!+\!p_{i2}^{\mu} ) \Bigg]
\nonumber \\
&&\hspace*{-8mm}- \Bigg[(p^{*}_f \cdot p_{f1}\!+\!p_{f2})\; (p_{i1}^{\mu} \!+\!p_{i2}^{\mu})
+(p_{i1}\!+\!p_{i2} \cdot p_{f1}\!+\!p_{f2})\;p^{*\,\mu}_f
- (p^{*}_f \cdot p_{i1}\!+\!p_{i2})\; (p_{f1}^{\mu} \!+\!p_{f2}^{\mu})
\nonumber \\
&&\hspace*{-8mm}\;\;\;+(p_{i2} \cdot p_{f1}\!+\!p_{f2})\;p^{\mu}_{i1}
-(p_{i1} \cdot p_{f1}\!+\!p_{f2})\;p^{\mu}_{i2}
+p_{i1}\!\cdot\! p_{i2}\;(p_{f1}^{\mu} \!+\!p_{f2}^{\mu})
+m^2\;(p_{f1}^{\mu} \!+\!p_{f2}^{\mu})\Bigg]\,.
\nonumber \\
\label{appc2}
\end{eqnarray}
In order to make the relation with the expression of the one-body current
contribution to the form factor, eq. (\ref{y3a}), the following points 
are noticed. The quark propagator can be written as the sum of two terms 
corresponding to ``positive" and ``negative" energies:  
\begin{equation}
\frac{\gamma \ccdot p^* +m_q}{m_q^2-p^{*\,2}}=
\frac{\xi^2\;(\gamma \ccdot \tilde{p}^{+} +m_q)}{2\,\xi \ccdot \tilde{p}^{+}\;
(\xi \ccdot \tilde{p}^{+}-\xi \ccdot p^*)}
+\frac{\xi^2\;(\gamma \ccdot \tilde{p}^{-} +m_q)}{2\,\xi \ccdot \tilde{p}^{-}\;
(\xi \ccdot \tilde{p}^{-}-\xi \ccdot p^*)}\, ,
\label{appc3}
\end{equation}
with
\begin{eqnarray}
&&\tilde{p}^{+\,\mu}=p^{*\,\mu}-\xi^{\mu}\;\Bigg(
\frac{\xi \ccdot p^* - \xi \ccdot \tilde{p} }{ 
\xi^2 }\Bigg) ,\;\;\;\;\;\;\;\;
\tilde{p}^{-\,\mu}=p^{*\,\mu}-\xi^{\mu}\;\Bigg(
\frac{\xi \ccdot p^* + \xi \ccdot \tilde{p} }{ 
\xi^2 }\Bigg)\, ,
\nonumber \\
&&\xi \ccdot \tilde{p}=
\sqrt{(\xi \ccdot p^*)^2+\xi^2\;(m_q^2\!-\!p^{*\,2}) }
=\xi \ccdot \tilde{p}^{+}=-\xi \ccdot \tilde{p}^{-},\;\;\;\;\;\;\;\;
\tilde{p}^{+\,2}=\tilde{p}^{-\,2}=m_q^2\, .
\label{appc4}
\end{eqnarray}
Of course, contributions corresponding to a ``negative" energy or 
to the part of the interaction with a pseudo-vector character, 
which cannot be accounted for in a RQM formalism and give rise 
to the two-body currents we looked for, are discarded at this point. 
Now, some replacements, which are justified by the lowest-order interaction 
terms of interest here, have to be made:
\begin{eqnarray}
&&\frac{g^2\;\frac{4}{3}}{\mu^2-(p_{i2}\!-\!p_{f2})^2} \hspace*{10mm}
\rightarrow 
\frac{g_{eff}^2\;\frac{4}{3}}{\mu^2+(\vec{k}_f-\vec{\tilde{k}}_f)^2}\, ,
\nonumber \\
&&\frac{\xi^2}{(\xi \ccdot \tilde{p}_f^{+}-\xi \ccdot p^*)} \hspace*{14mm}
\rightarrow 
\frac{2\,\xi \ccdot (\tilde{p}_f^{+}\!+\!p_{i2})}{
(\tilde{p}_f^{+}+p_{i2})^2-M^2 }\, ,
\nonumber \\
&&2\,\frac{(m_q^2\!+\!  p_{f1}\!\cdot\! p_{f2})}{m_q^2} \hspace*{10mm}
\rightarrow 
\frac{(p_{i2}\!+\!\tilde{p}^{+}_f)\ccdot(p_{f1} \!+\!p_{f2}) }{ 
m_q^2\!+\! \tilde{p}^{+}_f\ccdot p_{i2}}+
\Big(4\, \frac{e^2_{k_f}}{m_q^2}- 
\frac{e^2_{k_f}+e^2_{\tilde{k}_f}}{e^2_{\tilde{k}_f}}\Big)\, . 
\label{appc5}
\end{eqnarray}
Apart from the first term at the r.h.s. of the last expression, which 
will be included in two-body currents associated with a ``positive" 
energy intermediate state, the above replacements allow one to consider 
the integration over $\vec{p}_{f2}$ in eq. (\ref{appc1}). After adding 
the confining interaction, which has no direct role on the asymptotic behavior, 
one can use eq. (\ref{x2a} ) and then recovers the wave function 
$\tilde{\phi}(\tilde{k}_f)$. The identity with the expression 
of the form factor, $F_1^{(1)}(Q^2)$, eq. (\ref{y3a}), is finally obtained 
by making the change 
of notations: $p_{i2} \rightarrow p$, $p_{i1} \rightarrow p_i$, 
$\tilde{p}_f^{+}\rightarrow p_f$, $\tilde{k}_f\rightarrow k_f$.

The expression of two-body currents, which correspond to the terms discarded
above, can now be obtained:
\begin{eqnarray}
F^{(2)}_1(Q^2)= 
\int \frac{d\vec{p}_{i2}}{(2\,\pi)^3}\; 
\;\int \frac{d\vec{p}_{f2}}{(2\,\pi)^3}\;
\; \frac{\xi \ccdot (p_{i1}\!+\!p_{i2})}{2\,e_{i2}\;\;\xi \ccdot p_{i1}}
\;\Big(\frac{m_q}{e_{k_i}}\; \tilde{\phi}(k_i) \Big) 
\;\frac{\xi \ccdot (p_{f1}\!+\!p_{f2})}{2\,e_{f2}\;\;\xi \ccdot p_{f1}}
\;\Big(\frac{m_q}{e_{k_f}}\; \tilde{\phi}(k_f) \Big)
\nonumber \\ \times
\frac{g^2\;\frac{4}{3}}{\mu^2-(p_{i2}-p_{f2})^2}\;
\Big( I^{-}_{ps}+I^{-}_{pv}+ I^{+}_{ps}+I^{+}_{pv}\Big)
+ \;({\rm  term:} \;i \leftrightarrow f )\, , 
\label{twobodc1}
\end{eqnarray}
where:
\begin{eqnarray}
&&I^{-}_{ps}=\frac{\xi^2}{ \xi \ccdot (p_{i1}+p_{f1}+p_{i2}+p_{f2})\; \;  
2\,\xi \ccdot \tilde{p}_f\;(\xi \ccdot \tilde{p}_f+\xi \ccdot (P_f\!-\!p_{i2}) )}\;
\nonumber \\ 
&& \times\, 2\,\frac{m_q^2\!+\!  p_{f1}\!\cdot\! p_{f2}}{m_q^2}
\Bigg[  p_{i1} \ccdot p_{i2}\;\xi \ccdot \tilde{p}^{-}_f
 - p_{i1}\ccdot \tilde{p}^{-}_f\;\xi \ccdot p_{i2} 
 + p_{i2} \ccdot \tilde{p}^{-}_f\;\xi \ccdot p_{i1} 
 +m_q^2\,\xi \ccdot (\tilde{p}^{-}_f\! +\! p_{i1} \!+\!p_{i2} ) \Bigg]\, ,
\nonumber \\
&&I^{-}_{pv}=-\frac{\xi^2}{\xi \ccdot (p_{i1}+p_{f1}+p_{i2}+p_{f2}) \; \;  
2\,\xi \ccdot \tilde{p}_f\;(\xi \ccdot \tilde{p}_f+\xi \ccdot (P_f\!-\!p_{i2}) )}\;
\nonumber \\
&&\times \Bigg[
\tilde{p}^{-}_f \ccdot (p_{f1}\!+\!p_{f2})\; \xi \ccdot (p_{i1} \!+\!p_{i2})
+(p_{i1}\!+\!p_{i2}) \ccdot (p_{f1}\!+\!p_{f2})\;\xi \ccdot \tilde{p}^{-}_f
- \tilde{p}^{-}_f \ccdot (p_{i1}\!+\!p_{i2})\; \xi \ccdot (p_{f1} \!+\!p_{f2})
\nonumber \\
&&\;\;\;+p_{i2} \ccdot (p_{f1}\!+\!p_{f2})\; \xi \ccdot p_{i1}
-p_{i1} \ccdot (p_{f1}\!+\!p_{f2})\; \xi \ccdot p_{i2}
+(m_q^2+p_{i1}\ccdot p_{i2})\; \xi \ccdot (p_{f1} \!+\!p_{f2})
\Bigg]\,,
\nonumber \\
&&I^{+}_{ps}+I^{+}_{pv}=\frac{\xi^2}{ \xi \ccdot (p_{i1}+p_{f1}+p_{i2}+p_{f2})\; \;  
2\,\xi \ccdot \tilde{p}_f\;(\xi \ccdot \tilde{p}_f-\xi \ccdot (P_f\!-\!p_{i2}) )}\;
\nonumber \\  
&& \times \Bigg[
\frac{(p_{i2}\!+\!\tilde{p}^{+}_f)\ccdot(p_{f1} \!+\!p_{f2}) }{ 
m_q^2\!+\! \tilde{p}^{+}_f\ccdot p_{i2}}\,
\Big(   p_{i1} \ccdot p_{i2}\;\xi \ccdot \tilde{p}^{+}_f
 - p_{i1}\ccdot \tilde{p}^{+}_f\;\xi \ccdot p_{i2} 
 + p_{i2} \ccdot \tilde{p}^{+}_f\;\xi \ccdot p_{i1} 
 +m_q^2\,\xi \ccdot (\tilde{p}^{+}_f\! +\! p_{i1} \!+\!p_{i2} )\Big) 
\nonumber \\
&&- \Big(
\tilde{p}^{+}_f \ccdot (p_{f1}\!+\!p_{f2})\; \xi \ccdot (p_{i1} \!+\!p_{i2})
+(p_{i1}\!+\!p_{i2}) \ccdot (p_{f1}\!+\!p_{f2})\;\xi \ccdot \tilde{p}^{+}_f
- \tilde{p}^{+}_f \ccdot (p_{i1}\!+\!p_{i2})\; \xi \ccdot (p_{f1} \!+\!p_{f2})
\nonumber \\
&&\;\;\;+p_{i2} \ccdot (p_{f1}\!+\!p_{f2})\; \xi \ccdot p_{i1}
-p_{i1} \ccdot (p_{f1}\!+\!p_{f2})\; \xi \ccdot p_{i2}
+(m_q^2+p_{i1}\ccdot p_{i2})\; \xi \ccdot (p_{f1} \!+\!p_{f2})
\Big)\Bigg]
\nonumber \\
&&+\frac{\xi^2}{ 2\,\xi \ccdot \tilde{p}_f} \;
\Bigg[\frac{1}{ \xi \ccdot (p_{i1}+p_{f1}+p_{i2}+p_{f2})\; \; 
(\xi \ccdot \tilde{p}_f-\xi \ccdot (P_f\!-\!p_{i2}) }
\nonumber \\  && \hspace*{2cm}
-\frac{1}{ \xi \ccdot (p_{i1}+\tilde{p}_f+2\,p_{i2})}\; \; 
\Bigg(\frac{1}{\xi \ccdot (\tilde{p}_f+p_{i2})-\xi \ccdot P_f }
+\frac{1}{\xi \ccdot (\tilde{p}_f+p_{i2})+\xi \ccdot P_f }
\Bigg)\Bigg]
\nonumber \\  && \hspace*{1cm}\times
\Bigg[\Bigg( 2\,\frac{m_q^2\!+\!  p_{f1}\!\cdot\! p_{f2}}{m_q^2}
-\frac{(m_q^2\!+\!p_{i2}\ccdot\tilde{p}^{+}_f)) 
+(m_q^2\!+\!p_{f1} \ccdot p_{f2})}{m_q^2\!+\! \tilde{p}^{+}_f\ccdot p_{i2}} \Bigg)
\nonumber \\  && \hspace*{2cm} \times
\Big(   p_{i1} \ccdot p_{i2}\;\xi \ccdot \tilde{p}^{+}_f
 - p_{i1}\ccdot \tilde{p}^{+}_f\;\xi \ccdot p_{i2} 
 + p_{i2} \ccdot \tilde{p}^{+}_f\;\xi \ccdot p_{i1} 
 +m_q^2\,\xi \ccdot (\tilde{p}^{+}_f\! +\! p_{i1} \!+\!p_{i2} )\Big) \Bigg]\, .
\end{eqnarray}
Examination of the denominator of $I^-$ shows the presence of a factor, 
$\xi \ccdot \tilde{p}_f+\xi \ccdot (P_f\!-\!p_{i2})$, which has no support 
in time-ordered diagrams, as can be seen in schematic models with scalar 
particles. It is cancelled by another term with the result to be replaced 
by  $\tilde{p}_f+p_{i1}+\xi \ccdot (P_f\!-\!P_i)$, which amounts to
taking into account off-energy shell corrections. This has the advantage 
of removing a contribution that has a singular behavior in the limit 
of a massless pion (instant-form case). Thus, a more elaborate 
and less sensitive expression for $I^{-}_{ps}$ and $I^{-}_{pv}$,  
which we will use, is the following: 
\begin{eqnarray}
&&I^{-}_{ps}=\frac{\xi^2}{ \xi \ccdot (p_{i1}+p_{f1}+p_{i2}+p_{f2})\; \;  
2\,\xi \ccdot \tilde{p}_f\;\Big(\xi \ccdot (\tilde{p}_f+p_{i1})+
\xi \ccdot (P_f\!-\!P_i) \Big)}\;
\nonumber \\ 
&& \times 2\,\frac{m_q^2\!+\!  p_{f1}\!\cdot\! p_{f2}}{m_q^2}
\Bigg[  p_{i1} \ccdot p_{i2}\;\xi \ccdot \tilde{p}^{-}_f
 - p_{i1}\ccdot \tilde{p}^{-}_f\;\xi \ccdot p_{i2} 
 + p_{i2} \ccdot \tilde{p}^{-}_f\;\xi \ccdot p_{i1} 
 +m_q^2\,\xi \ccdot (\tilde{p}^{-}_f\! +\! p_{i1} \!+\!p_{i2} ) \Bigg]\, ,
\nonumber \\
&&I^{-}_{pv}=-\frac{\xi^2}{\xi \ccdot (p_{i1}+p_{f1}+p_{i2}+p_{f2}) \; \;  
2\,\xi \ccdot \tilde{p}_f\;(\xi \ccdot (\tilde{p}_f+p_{i1} )
+\xi \ccdot (P_f\!-\!P_i) )}\;
\nonumber \\
&&\times \Bigg[
\tilde{p}^{-}_f \ccdot (p_{f1}\!+\!p_{f2})\; \xi \ccdot (p_{i1} \!+\!p_{i2})
+(p_{i1}\!+\!p_{i2}) \ccdot (p_{f1}\!+\!p_{f2})\;\xi \ccdot \tilde{p}^{-}_f
- \tilde{p}^{-}_f \ccdot (p_{i1}\!+\!p_{i2})\; \xi \ccdot (p_{f1} \!+\!p_{f2})
\nonumber \\
&&\;\;\;+p_{i2} \ccdot (p_{f1}\!+\!p_{f2})\; \xi \ccdot p_{i1}
-p_{i1} \ccdot (p_{f1}\!+\!p_{f2})\; \xi \ccdot p_{i2}
+(m_q^2+p_{i1}\ccdot p_{i2})\; \xi \ccdot (p_{f1} \!+\!p_{f2})
\Bigg]\,.
\end{eqnarray}
In the instant-form approach, and for the Breit frame, the above contribution 
does not show any explicit dependence on the pion mass. In the front-form 
approach, it vanishes due to the presence of the factor $\xi^2$. 

The expression for $I^{+}_{ps}+I^{+}_{pv}$ can also be transformed. The factor 
at the denominator, $\xi \ccdot \tilde{p}_f-\xi \ccdot (P_f\!-\!p_{i2})$, 
can be replaced by $\xi \ccdot (\tilde{p}_f+p_{i2}-p_{f1}-p_{f2})$, 
which again amounts to neglect off-energy shell corrections. This term 
may be more singular but it turns out that the numerator vanishes 
at the same time. Some simplification can therefore be made. Moreover, 
the whole quantity goes to zero with the momentum transfer, allowing to
factorize  the term $\tilde{p}^{+}_f\!-\!p_{i1}$. The resulting
expression we will use is thus:
\begin{eqnarray}
I^{+}_{ps}+I^{+}_{pv} &=&-\frac{1}{ \xi \ccdot (p_{i1}+p_{f1}+p_{i2}+p_{f2})\; \;  
2\,\xi \ccdot \tilde{p}_f\; (m_q^2\!+\! \tilde{p}^{+}_f\ccdot p_{i2})}\;
\nonumber \\
&&\hspace*{-5mm} \times \Bigg[
\xi \ccdot \tilde{p}^{+}_f\;\;\Big((\tilde{p}^{+}_f\!-\!p_{i1}) \ccdot  \xi
\;\;(m_q^2\!+\! p_{i1}\ccdot p_{i2})
-(p_{i1}\!+\!p_{i2}) \ccdot  \xi \;\;(\tilde{p}^{+}_f\!-\!p_{i1}) \ccdot p_{i2}\Big)
\nonumber \\
&&-(\tilde{p}^{+}_f\!+\!p_{i2}) \ccdot  \xi \;\; p_{i2} \ccdot  \xi \;\;
(\tilde{p}^{+}_f\!-\!p_{i1}) \ccdot p_{i1}
\nonumber \\
&& + \Big((\tilde{p}^{+}_f\!-\!p_{i1}) \ccdot (p_{i1}\!+\!p_{i2})\; \xi^2
- (\tilde{p}^{+}_f\!-\!p_{i1}) \ccdot \xi \;\; p_{i2} \ccdot  \xi \Big)
\; (m_q^2\!+\! \tilde{p}^{+}_f\ccdot p_{i2})\Bigg]
\nonumber \\
&& \hspace*{-1cm}+\frac{\xi^2 \;(\tilde{p}^{+}_f\!-\!p_{i1}) \ccdot  \xi 
}{ \xi \ccdot (p_{i1}+p_{f1}+p_{i2}+p_{f2})\; \; 2\,\xi \ccdot \tilde{p}_f \;
\xi \ccdot (p_{i1}+\tilde{p}_f+2\,p_{i2})\; 
\xi \ccdot (p_{i2}+\tilde{p}_f+p_{f1}+p_{f2})}\nonumber \\
&& \times 
\Bigg[\Bigg( 2\,\frac{m_q^2\!+\!  p_{f1}\!\cdot\! p_{f2}}{m_q^2}
-\frac{(m_q^2\!+\!p_{i2}\ccdot\tilde{p}^{+}_f)) 
+(m_q^2\!+\!p_{f1} \ccdot p_{f2})}{m_q^2\!+\! \tilde{p}^{+}_f\ccdot p_{i2}} \Bigg)
\nonumber \\  && \hspace*{1cm} \times
\Big(   p_{i1} \ccdot p_{i2}\;\xi \ccdot \tilde{p}^{+}_f
 - p_{i1}\ccdot \tilde{p}^{+}_f\;\xi \ccdot p_{i2} 
 + p_{i2} \ccdot \tilde{p}^{+}_f\;\xi \ccdot p_{i1} 
 +m_q^2\,\xi \ccdot (\tilde{p}^{+}_f\! +\! p_{i1} \!+\!p_{i2} )\Big) \Bigg]\, .
\nonumber \\
\end{eqnarray}
It can be checked that the above expression vanishes for both the instant form 
and the Breit frame ($\vec{P}_f=-\vec{P}_i$). It does not however away 
from this configuration. In the configuration $E_i=E_f$ together with the limit 
$|\vec{P}_i+\vec{P}_i| \rightarrow \infty$, the expression is found to be
identical to the usual front-form one ($q^+=0$). This one could be obtained
from the above one using standard changes of variable.

The derivation of two-body currents considered in this appendix could be easily
extended to the Lorentz-scalar form factor. As this one already evidences 
the expected asymptotic power law, except perhaps for a numerical factor, 
the need for considering two-body currents is not as strong as for 
the charge form factor. Moreover, a perturbative calculation may reveal 
to provide little effect in the $Q^2$ domain where it does for the other form
factor. 


\end{document}